\documentclass[
    pra,
    aps,
    reprint,
    twocolumn,
    superscriptaddress,
    amsmath,
    amssymb,
    nofootinbib
]{revtex4-2}

\usepackage{graphicx}
\usepackage{dcolumn}
\usepackage{bm}
\usepackage{physics}
\usepackage{hyperref}
\usepackage{xcolor}
\usepackage{subfigure}
\usepackage[T1]{fontenc}
\newcommand{\orcid}[1]{%
  \href{https://orcid.org/#1}{\,\protect\includegraphics[width=7pt]{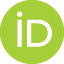}}%
}
\graphicspath{{Figures/}{}}

\begin{document}

\title{Geometric Control of Cat States in High Harmonic Generation}

\author{Arti Gaharwar\orcid{0009-0005-5820-0472}}
\email[]{Arti.Gaharwar@icfo.eu}
\affiliation{ICFO -- Institut de Ciencies Fotoniques, The Barcelona Institute of Science and Technology, 08860 Castelldefels (Barcelona), Spain}

\author{Rocío Borrego-Varillas\orcid{0000-0002-4499-0558}}
\affiliation{Istituto di Fotonica e Nanotecnologie, Consiglio Nazionale delle Ricerche, Milan, Italy}

\author{Marcelo F. Ciappina\orcid{0000-0002-1123-6460}}
\affiliation{Department of Physics, Guangdong Technion - Israel Institute of Technology,
241 Daxue Road, Shantou, Guangdong, China, 515063}
\affiliation{Technion – Israel Institute of Technology, Haifa, 32000, Israel}
\affiliation{Guangdong Provincial Key Laboratory of Materials and Technologies for Energy Conversion,
Guangdong Technion – Israel Institute of Technology, 241 Daxue Road, Shantou, Guangdong, China, 515063}

\author{Anna G. Ciriolo}
\affiliation{Istituto di Fotonica e Nanotecnologie, Consiglio Nazionale delle Ricerche, Milan, Italy}

\author{Javier Rivera-Dean}
\affiliation{Department of Physics and Astronomy, University College London, Gower Street, London WC1E 6BT, UK}

\author{Philipp Stammer}
\affiliation{ICFO -- Institut de Ciencies Fotoniques, The Barcelona Institute of Science and Technology, 08860 Castelldefels (Barcelona), Spain}

\author{Paraskevas Tzallas}
\affiliation{Foundation for Research and Technology-Hellas, Institute of Electronic Structure \& Laser, GR-70013 Heraklion (Crete), Greece}
\affiliation{Center for Quantum Science and Technologies (FORTH-QuTech), GR-70013 Heraklion (Crete), Greece}
\affiliation{ELI-ALPS, ELI-Hu Non-Profit Ltd., Dugonics tér 13, H-6720 Szeged, Hungary}

\author{Emilio Pisanty}
\affiliation{Attosecond Quantum Physics Laboratory, Department of Physics, King’s College London, Strand Campus, London WC2R 2LS, United Kingdom}

\author{Maciej Lewenstein\orcid{0000-0002-0210-7800}}
\email[]{maciej.lewenstein@icfo.eu}
\affiliation{ICFO -- Institut de Ciencies Fotoniques, The Barcelona Institute of Science and Technology, 08860 Castelldefels (Barcelona), Spain}

\begin{abstract}
High-harmonic generation (HHG) provides a powerful platform for exploring the interaction between intense laser fields and matter on ultrafast timescales. Beyond its conventional description in terms of emitted radiation and electron dynamics, a fully quantum treatment of HHG reveals that the nonlinear light–matter interaction can modify the quantum state of the driving field itself, establishing correlations between the fundamental and harmonic modes. This perspective opens new possibilities for using HHG as a tool to engineer and control nonclassical states of light. In this work, we investigate the geometric properties of optical Schrödinger cat states generated in HHG via conditioning and post-selection. By analyzing the coherent-state displacements induced by different structured driving fields, we characterize the resulting phase-space evolution and the associated geometric phases of the generated quantum states. Particular emphasis is placed on how the polarization and spatial-mode structure of the driving light influence the geometry and evolution of displaced coherent states. Our results demonstrate that structured light offers a versatile means of controlling the geometric dynamics of HHG-generated cat states across the parameter space. Furthermore, we discuss the prospects for realizing genuinely topological optical cat states by exploiting more complex structured light configurations.
\end{abstract}

\maketitle

\section{Introduction}
High-harmonic generation (HHG) is the key strong-field process that occurs at laser intensities of $10^{13}-10^{15}\,\mathrm{W/cm^2}$, where the laser electric field becomes comparable to the atomic Coulomb field. In this regime, the interaction between an atom and an intense laser field leads to the emission of radiation at integer multiples of the laser's frequency. The main features of HHG spectra are well explained by semiclassical models such as the three-step model \cite{corkum1993plasma,kulander_dynamics_1993}, and its quantum version, the strong-field approximation (SFA)  \cite{lewenstein1994theory}; these approaches capture the essential electron dynamics in the presence of a {\it classical driving field}. Therefore, since in these models the electromagnetic (EM) field is treated classically, its quantum state  remains unchanged during the interaction.

A fully quantized approach to HHG \cite{gonoskov2016quantum,lewenstein2021generation,stammer2023quantum,rivera2024nonclassical,stammer2025colloquium,yi2025generation,lange2024electron,lange2025hierarchy,stammer2024entanglement,gorlach2020quantum,lange2026high, stammer2026high} goes further by treating matter and radiation quantum mechanically. In this framework, harmonic emission can be understood as a redistribution of photons among quantized radiation modes, thereby enabling the study of photon statistics and correlations beyond spectral intensity measurements \cite{gonoskov2016quantum}. For a strongly coherent driving field, a semiclassical description is often sufficient, and quantum fluctuations are typically neglected \cite{glauber1963coherent,Walls&Milburn}. However, recent studies have shown that this approximation can fail in HHG, where quantizing the electromagnetic field becomes essential.
\begin{itemize}
\item A conditioning on HHG, combined with post-selection, monitoring changes of the driving field due to interactions with the target, leads to the generation of massively quantum states (MQS), optical Schrödinger cats. hese states can be described either within a simplified effective framework, as introduced in \cite{lewenstein2021generation}, or using the rigorous, obviously more complex approach that incorporates energy conservation, which is responsible for the buildup of quantum correlations \cite{rivera2024quantum,Stammer_EnergyConservation2024,SRL22,RLP22}.

\item Nonlinear interaction between atoms and strong laser fields can generate non-classical radiation and build correlations between the fundamental driving mode and the emitted harmonic modes, typically leading to the generation of MQS in the form of multimode squeezed states \cite{stammer2024entanglement,Stammer_EnergyConservation2024}. Related experimental studies have investigated quantum-light-driven HHG and the generation of nonclassical states of light \cite{TCS24}. 

\item Finally, quantum light can be used from the outset, for example in the form of bright squeezed vacuum (BSV) or bright squeezed light (BSL)\cite{GTB22,Maria_ArXiv_2024,Ido-Nirit-new, stammer2024limitations}. This research direction has also been further developed in subsequent studies \cite{RSC24,paris-BSL-prop,Lidija-1-mode,Javier-ATI-BSL,Gauss-Technion, stammer2026attosecond, stammer2024absence, rivera2026attosecond, stammer2026fluctuation}.
\end{itemize}

The above theoretical predictions and, more importantly, the first experimental observations, open the way toward fascinating applications of quantum electrodynamics (QED) and quantum optics (QO) of HHG. These include, to name a few: nonlinear optics \cite{lamprou2025nonlinear}, quantum metrology \cite{stammer2024metrological, stammer2026attosecond}, quantum optical coherence studies and antibunching~\cite{stammer2025theory,stammer2026quantum,stammer2026photon}, weak measurements \cite{SRC25} and many more.

An important consequence of the quantum-optical description of HHG is that the quantum state of the driving laser field is modified during the strong-field interaction. As photons are transferred from the fundamental driving mode to the emitted harmonic modes, the distribution of photons across the modes is correlated because of energy conservation \cite{rivera2024quantum,Stammer_EnergyConservation2024}. 
Post-selection projects the final field state onto the component of the fundamental mode that has been modified by the HHG process. This produces a coherent superposition of the initial and amplitude-shifted coherent states  \cite{lewenstein2021generation}. The resulting superposition forms an optical Schrödinger cat state \cite{schrodinger1935gegenwartige,yurke1986generating}. Beyond atomic HHG, optical Schrödinger cat states have also been proposed in other strong-field platforms, including above-threshold ionization and semiconductor systems \cite{RLP22,gan2026tailoring}.

Building on this progress, recent studies have shown that optical Schrödinger cat states generated via HHG exhibit significantly greater robustness to photon loss than conventional even and odd cat states, while retaining their metrological advantage under realistic lossy conditions (cf. \cite{stammer2024metrological}). These properties make them promising candidates for quantum metrology and other quantum information technologies. These advances open new opportunities to investigate how the properties of optical Schrödinger cat states can be engineered and controlled.

A natural route to such control is provided by structured driving fields, which have already been widely employed in HHG to manipulate the angular momentum, polarization, and spatial structure of the emitted harmonics~\cite{gan2026tailoring,forbes2019structured, das2026optical}. In particular, beams carrying orbital angular momentum (OAM), vector beams, and Full Poincaré beams (FPBs), also referred to in recent literature as optical Stokes skyrmions~\cite{Gao2020}, provide versatile control over the properties of high-harmonic radiation~\cite{USVortex,das2026optical,allen1999iv,forbes2019structured,Beckley2010,milione2011higher,kong2017controlling,pisanty2019conservation,paufler2019high,bai2025dynamical,cisowski2026geometric}. The additional degrees of freedom offered by these structured driving fields therefore provide a natural set of externally tunable control parameters for engineering the geometric and topological properties of optical Schrödinger cat states.

In this work, we investigate the geometric and topological properties of optical Schrödinger cat states generated through HHG in gas media. Our aim is to understand how the coherent-state components forming these optical cat states evolve as the control parameters of structured driving fields are varied. We consider two structured driving configurations. The first one employs a family of rotating polarization ellipses with fixed ellipticity, allowing continuous control of the polarization ellipse's orientation. The second approach uses Full Poincaré beams carrying OAM, whose spatially varying phase and polarization distributions provide an additional degree of control. By varying these driving-field parameters, we steer the trajectories of the coherent-state components in parameter space, thereby controlling the geometry of the optical cat states and the geometric phases they acquire. In this paper, we use the approach as in Ref.~\cite{lewenstein2021generation}, encouraged by the fact that in the case of single mode driving with linear polarization, this approach gives a better fidelity of the final state than 97\% (see \cite{rivera2024quantum}). Our results demonstrate that structured strong-field driving provides a versatile means of controlling the geometry and geometric phases of optical Schrödinger cat states.

\section{Theoretical Framework}

\subsection{Optical Cat States in Intense Laser–Matter Interactions}
This section introduces the theoretical description of optical Schrödinger cat states generated in HHG. We establish the general form of the conditional cat state and define the interaction-induced coherent-state displacements, which constitute the key quantities for the geometric and topological analysis presented in the following sections. In this paper, we use the description of the cat-state generation presented in \cite{lewenstein2021generation}, motivated by the rigorous analyses reported in \cite{rivera2024quantum,Stammer_EnergyConservation2024}.

In the fully quantized description of HHG, the interaction between an atom and the driving laser field modifies the quantum state of the fundamental mode and determines the quantum states of harmonic modes. When the (linearly polarized) radiation field is conditioned on the detection of harmonic photons and on changes of the driving field due to interactions \cite{rivera2024quantum}, the quantum state of the fundamental mode, to a very good approximation,  is projected onto a superposition of coherent states. The resulting post-selected state of the fundamental mode can be written in a simplified, but accurate form as \cite{lewenstein2021generation,rivera2024quantum}
\begin{equation}
|\psi_{\mathrm{cond}}\rangle
=
\frac{1}{N}
\left(
|\alpha+\Delta\alpha\rangle
-
\zeta\,|\alpha\rangle
\right),
\label{cat}
\end{equation}
where $|\alpha\rangle$ denotes the initial coherent state of the driving field, and $\Delta\alpha$ represents the displacement of the coherent amplitude induced by the nonlinear interactions. In Eq.~(\ref{cat}), the overlap between the initial and displaced coherent states is given by
\begin{equation}
\zeta = \langle \alpha|\alpha+\Delta\alpha\rangle,
\label{overlap}
\end{equation}
and the normalization constant is
\begin{equation}
N = \sqrt{1-|\zeta|^2}.
\end{equation}
For a driving field consisting of two quantized driving field modes, the conditional state becomes~\cite{SRL22,stammer2022theory}
\begin{equation}
|\psi_{\mathrm{cond}}\rangle
=
\frac{1}{N}
\left(
|\alpha+\Delta\alpha\rangle \otimes |\beta+\Delta\beta\rangle
-
\zeta_{\alpha}\zeta_{\beta}\,
|\alpha\rangle \otimes |\beta\rangle
\right),\label{eq4}
\end{equation}
where $\alpha$ and $\beta$ denote the initial coherent amplitudes of the two driving field modes of the same carrier frequency, while $\Delta\alpha$ and $\Delta\beta$ represent the interaction-induced displacements of the corresponding coherent states. As a consequence of the atom–field interaction, the amplitudes of both modes undergo shifts, leading to nonzero overlaps between the initial and displaced coherent states. These overlap factors are incorporated through the quantities $\zeta_\alpha$ and $\zeta_\beta$.

The coherent-state displacements are governed by the laser-driven atomic dipole response. Let us assume that the two-mode driving field is characterized by two control parameters $\boldsymbol\lambda=(\lambda_1,\lambda_2)$; the specific physical interpretation of these parameters depends on the structure of the driving field being considered. The interaction-induced coherent-state displacements can then be formally written as follows:
\begin{equation}
\Delta\alpha(\boldsymbol\lambda)
=
g_{\alpha}
\int_{0}^{t} dt'\,
f_{\alpha}(t')\,
\langle \mathbf{d}\cdot\boldsymbol{\epsilon}_{\alpha} \rangle
(t';\boldsymbol\lambda)\,
e^{i\omega_{\alpha} t'},
\label{delta_alpha}
\end{equation}
and
\begin{equation}
\Delta\beta(\boldsymbol\lambda)
=
g_{\beta}
\int_{0}^{t} dt'\,
f_{\beta}(t')\,
\langle \mathbf{d}\cdot\boldsymbol{\epsilon}_{\beta} \rangle
(t';\boldsymbol\lambda)\,
e^{i\omega_{\beta} t'}.
\label{delta_beta}
\end{equation}

This explicit dependence on the control parameters, introduced in Eqs.~(\ref{delta_alpha}) and (\ref{delta_beta}), is the key element of our analysis, since it defines the paths explored in the parameter space $(\boldsymbol\lambda=\lambda_1,\lambda_2)$, and consequently determines the associated geometric phases. The coupling constants $g_{\alpha}$ and $g_{\beta}$ quantify the interaction between the atomic dipole and the two quantized driving modes, while $f_{\alpha}(t)$ and $f_{\beta}(t)$ describe the temporal profiles of the corresponding pulses. The vectors $\boldsymbol{\epsilon}_{\alpha}$ and $\boldsymbol{\epsilon}_{\beta}$ specify the polarization directions of the two modes. The laser-driven dipole response $\langle \mathbf{d}\cdot\boldsymbol{\epsilon}_{\beta} \rangle
(t';\boldsymbol\lambda)$ carries the imprint of the structured driving field through its dependence on the laser electric field $\mathbf{E}(t;\boldsymbol\lambda)$ and its associated vector potential $\mathbf{A}(t;\boldsymbol\lambda)$, which determine the strong-field evolution of the atomic system.
Within the strong-field approximation (SFA), the time-dependent dipole moment can be expressed as \cite{lewenstein1994theory,smirnova2014multielectron}
\begin{equation}
\begin{aligned}
\left\langle \mathbf{d}\cdot\boldsymbol{\epsilon}_j \right\rangle
(t;\boldsymbol\lambda)
&=
{\rm Re}\Bigg\{
i\int_{0}^{t} dt'
\int d^{3}\mathbf{p}\;
\boldsymbol{\epsilon}_j\cdot
\mathbf{d}_{g}^{*}\!\left(\mathbf{p}-\mathbf{A}(t;\boldsymbol\lambda)\right)
\\
&\hspace{-1.0cm}\times
\exp\!\left[-iS(t,t',\mathbf{p};\boldsymbol\lambda)\right]
\Big[
\mathbf{d}_{g}\!\left(\mathbf{p}-\mathbf{A}(t';\boldsymbol\lambda)\right)
\cdot
\mathbf{E}(t';\boldsymbol\lambda)
\Big]
\Bigg\}.
\end{aligned}
\label{eq7}
\end{equation}
where \(j=\alpha,\beta\) and $\mathbf{d}_{g}(\mathbf{k})=\langle k|\mathbf{d}|\mathbf{g}\rangle$ denotes the transition dipole matrix element between the ground state $|g\rangle$ and an electron  state of momentum $\mathbf{k}$ in the continuum. The quasi-classical action appearing in the phase factor is given by
\begin{equation}
S(t,t',\mathbf{p};\boldsymbol\lambda)
=
\frac{1}{2}\int_{t'}^{t} d\tau\,
\big[\mathbf{p}-\mathbf{A}(\tau;\boldsymbol\lambda)\big]^2
+
I_p (t-t'),
\end{equation}
with $I_p$ denoting the ionization potential. The dependence of the displacements $\Delta\alpha(\boldsymbol\lambda)$ and $\Delta\beta(\boldsymbol\lambda)$ on the driving-field parameters $\boldsymbol\lambda=(\lambda_1,\lambda_2)$ arises from the electric field and vector potential that govern the strong-field electron dynamics. These fields determine the semiclassical electron trajectories and, through the induced atomic dipole response, define the resulting coherent-state displacements.

\subsection{Family of Rotating Polarization Ellipses (FRE)}

In this section, we introduce the family of rotating polarization ellipses (FRE) as the first structured driving-field configuration considered in this work. We define the corresponding parameter-dependent driving field and the associated conditional optical Schrödinger cat states, which provide the basis for the geometric analysis presented in the following sections.

A controlled parameter space for investigating geometric effects is constructed by
considering a two-mode driving field composed of two orthogonal linear
polarization modes in the transverse $xy$-plane. The corresponding classical
electric field is written as
\begin{equation}
\mathbf{E}(t;\theta,\phi)
=
E_x(t;\theta,\phi)\,\hat{\mathbf{x}}
+
E_y(t;\theta,\phi)\,\hat{\mathbf{y}}.
\label{eq9}
\end{equation}
The field components in the quadrature form are expressed as
\begin{equation}
\begin{pmatrix}
E_x \\ 
E_y
\end{pmatrix}
=
E_0 f(t)
\Big[
\cos(\omega t)\,\mathbf{u}(\theta,\phi)
+
\sin(\omega t)\,\mathbf{v}(\theta,\phi)
\Big],
\end{equation}
where
\begin{align}
\mathbf{u}(\theta,\phi)
&=
\begin{pmatrix}
\cos\theta \cos\phi \\
\cos\theta \sin\phi
\end{pmatrix},
\\
\mathbf{v}(\theta,\phi)
&=
\begin{pmatrix}
\sin\theta \sin\phi \\
-\sin\theta \cos\phi
\end{pmatrix}.
\end{align}

Here, $E_0$ denotes the peak electric-field amplitude, $\omega$ is the carrier
frequency, and $f(t)$ represents a real pulse envelope (e.g.,
$f(t)=\sin^2(\frac{\omega t}{2n_{\mathrm{cyc}}})$ over a finite pulse duration, where $n_{\mathrm{cyc}}$ is the total number of cycles). The parameters
$(\theta,\phi)$ define a two-dimensional parameter space describing the
polarization state of the driving field \cite{bhandari1997polarization}. The
parameter $\theta$ controls the ellipticity of the polarization ellipse, whereas
$\phi$ determines its orientation in the transverse plane. By fixing $\theta$ and
varying $\phi$, one obtains a family of rotating polarization ellipses with
constant ellipticity (see Fig.~\ref{fig:structured_fields}(a) in the Supplementary Material).

The corresponding quantum description is obtained by associating each
polarization component with a coherent-state amplitude. The initial state of the radiation field is therefore written as
\begin{equation}
|\Psi_0(\theta,\phi)\rangle
=
|\alpha_x(\theta,\phi)\rangle
\otimes
|\alpha_y(\theta,\phi)\rangle ,
\label{eq13}
\end{equation}
where the coherent-state amplitudes are given by
\begin{equation}
\alpha_x(\theta,\phi)
=
\alpha_0
\left[
\cos\theta \cos\phi
-
i \sin\theta \sin\phi
\right]
\label{eq14}
\end{equation}
and
\begin{equation}
\alpha_y(\theta,\phi)
=
\alpha_0
\left[
\cos\theta \sin\phi
+
i \sin\theta \cos\phi
\right].
\label{eq15}
\end{equation}

The parameter $\alpha_0$ sets the overall coherent-state amplitude and
determines the mean photon-number $N = |\alpha_0|^2$. The interaction-induced
coherent-state displacements are denoted by
$\Delta\alpha_x(\theta,\phi)$ and $\Delta\alpha_y(\theta,\phi)$, corresponding
to the initial coherent amplitudes $\alpha_x(\theta,\phi)$ and
$\alpha_y(\theta,\phi)$ defined in Eqs.~(\ref{eq13})--(\ref{eq15}). These
displacements describe the quantum back-action of the HHG process on the two
polarization modes. They are obtained by evaluating the SFA dipole expressions
of Eqs.~(\ref{delta_alpha})--(\ref{eq7}) for the driving electric field
$E(t;\theta,\phi)$, defined in Eq.~(\ref{eq9}), which generates the family of
rotating polarization ellipses.

Using these mode-resolved displacements, the HHG-conditioned cat state, generated
by the FRE driving field, can be expressed as a two-mode coherent state
superposition, as in Eqs.~(\ref{eq4}):
\begin{eqnarray}
&&|\psi_{\mathrm{FRE}}(\theta,\phi)\rangle
=
\frac{1}{\mathcal{N}}
\left[
|\alpha_x(\theta,\phi)+\Delta\alpha_x(\theta,\phi)\rangle
\otimes
|\alpha_y(\theta,\phi)\right.\nonumber \\&&\left.+\Delta\alpha_y(\theta,\phi)\rangle
-\zeta_x\zeta_y
|\alpha_x(\theta,\phi)\rangle
\otimes
|\alpha_y(\theta,\phi)\rangle
\right],
\label{eq:FRE_cat}
\end{eqnarray}
where $\zeta_x=\langle\alpha_x|\alpha_x+\Delta\alpha_x\rangle$ and
$\zeta_y=\langle\alpha_y|\alpha_y+\Delta\alpha_y\rangle$ quantify the overlaps
between the initial and displaced coherent states of the two polarization
modes. Since both coherent amplitudes and  HHG-induced displacements
depend explicitly on $(\theta,\phi)$, the conditional cat state defines a
parameter-dependent family of quantum states.

This parametrization provides a natural two-dimensional control manifold, in
which the driving field can evolve along closed loops. The resulting
parameter-dependent coherent-state displacements enable the exploration of the
geometric structure and associated geometric phases of the HHG-generated
conditional cat states.

\subsection{Full Poincaré Beam (FPB)}
In this subsection, we introduce the Full Poincaré Beam (FPB) as the second structured driving-field configuration considered in this work. We define its spatial mode structure and control parameters, and describe the corresponding HHG-induced coherent-state displacements and conditional optical Schrödinger cat state.

To extend the analysis beyond polarization-only control, we consider spatially structured driving fields in the form of a Full Poincar\'e Beam (FPB)~\cite{forbes2019structured}, which introduces additional spatial degrees of freedom. An FPB is generated by the coherent superposition of a
Gaussian mode and a Laguerre--Gaussian vortex mode carrying opposite circular polarizations, resulting in a spatially varying polarization distribution across the transverse beam profile. This additional spatial structure provides an extra degree of control over the light--matter interaction.
The corresponding classical electric field is written as
\begin{equation}
\mathbf E(\mathbf r,t)
=
\Re\!\left\{
e^{-i\omega t}
\Big[
\alpha_g U_g(\rho)\,\mathbf e_{+}
+
\alpha_v U_v(\rho,\varphi)\,\mathbf e_{-}
\Big]
\right\},
\label{eq17}
\end{equation}
where $\mathbf e_{\pm}$ denote the circular polarization unit vectors, with $\mathbf e_{+}$ associated with the Gaussian mode and $\mathbf e_{-}$ with the vortex mode. The transverse spatial modes satisfy the normalization condition
\begin{equation}
\int_0^\infty \rho\, d\rho
\int_0^{2\pi} d\varphi\,
|U_j(\rho,\varphi)|^2
=
1,
\qquad
j=g,v.
\end{equation}

The Gaussian mode is described by
\begin{equation}
U_g(\rho)
=
\sqrt{\frac{2}{\pi}}
\frac{1}{w_0}
\exp\!\left(-\frac{\rho^2}{w_0^2}\right),
\end{equation}
whereas the Laguerre--Gaussian vortex mode is given
by~\cite{allen1999iv}
\begin{equation}
U_v(\rho,\varphi)
=
\sqrt{\frac{2}{\pi|\ell|!}}
\frac{1}{w_0}
\left(
\frac{\sqrt2\,\rho}{w_0}
\right)^{|\ell|}
\exp\!\left(-\frac{\rho^2}{w_0^2}\right)
e^{i\ell\varphi},
\end{equation}
where $\ell$ denotes the orbital angular momentum (OAM) charge, $\mathbf r=(\rho,\varphi)$ is the transverse position in polar coordinates, and $w_0$ is the beam waist of the transverse modes. The Gaussian mode is localized around the beam-center, whereas the Laguerre--Gaussian mode carries OAM and exhibits a phase singularity at the origin~\cite{yao2011orbital,das2026optical}. Throughout this work, transverse mode profiles are evaluated at a fixed reference plane
($z=0$). Macroscopic propagation effects, including diffraction, Gouy phase accumulation, and phase matching, are neglected, corresponding to the thin-medium approximation~\cite{das2026optical}.

The driving field is characterized by coherent-state amplitudes
\begin{equation}
\alpha_g=\alpha_0\cos\theta,
\qquad
\alpha_v=\alpha_0\sin\theta\,e^{i\delta},
\label{FPB_amplitude}
\end{equation}
where $\theta$ controls the relative intensity between the Gaussian and vortex modes, while $\delta$ defines their relative optical phase. Together, these parameters span a two-dimensional control space for the structured driving field.
The initial state of the driving field is therefore written as
\begin{equation}
|\Psi_0(\theta,\delta)\rangle
=
|\alpha_g(\theta,\delta)\rangle_g
\otimes
|\alpha_v(\theta,\delta)\rangle_v.
\end{equation}

The HHG process produces a spatially dependent dipole response, resulting in local coherent-state displacements in the circular polarization basis,
\begin{equation}
\Delta\alpha_{\pm}(\rho,\varphi;\lambda)
=
g
\int dt\,
D_{\pm}(t;\rho,\varphi,\lambda)
e^{i\omega t},
\label{23}
\end{equation}
where $D_{\pm}$ denote the dipole components in the circular basis and $\lambda$ represents an external control parameter of the structured driving field (e.g., $\theta$ or $\delta$) and $g$ is the light–matter coupling constant. The experimentally accessible displacements of the detected transverse modes are obtained by projecting the local displacements onto the corresponding spatial mode functions,
\begin{align}
\Delta\alpha_g(\lambda)
&=
n_A
\int_0^\infty
\rho\,d\rho
\int_0^{2\pi}
d\varphi\,
U_g^*(\rho)\,
\Delta\alpha_{+}(\rho,\varphi;\lambda),\label{24}
\\
\Delta\alpha_v(\lambda)
&=
n_A
\int_0^\infty
\rho\,d\rho
\int_0^{2\pi}
d\varphi\,
U_v^*(\rho,\varphi)\,
\Delta\alpha_{-}(\rho,\varphi;\lambda).
\label{25}
\end{align}

Here, $n_A$ denotes the areal density of emitters within the interaction region and accounts for the coherent contribution of the ensemble of atoms distributed across the transverse plane. Consequently, the corresponding two-mode conditional optical cat state can be expressed as
\begin{eqnarray}
&&|\psi_{\mathrm{FPB}}(\theta,\delta)\rangle
=
\frac{1}{N}
\left[
|\alpha_g(\theta,\delta)+\Delta\alpha_g(\theta,\delta)\rangle
\otimes
|\alpha_v(\theta,\delta)\right. \nonumber \\
&&\left.+\Delta\alpha_v(\theta,\delta)\rangle
-
\zeta_g\zeta_v
|\alpha_g(\theta,\delta)\rangle
\otimes
|\alpha_v(\theta,\delta)\rangle
\right].
\label{eq:FPB_cat}
\end{eqnarray}

\section{Parameter Manifold and Geometric Phase}

The conditional optical cat states introduced in
Eqs.~(\ref{eq:FRE_cat}) and~(\ref{eq:FPB_cat}) depend explicitly on the external
control parameters through both the initial coherent-state amplitudes and the
interaction-induced HHG displacements. For the family of rotating polarization
ellipses (FRE), the control manifold is defined by
$\lambda=(\theta,\phi)$, whereas for the full Poincaré beam (FPB) it is given by
$\lambda=(\theta,\delta)$. As these parameters are continuously varied, the
corresponding conditional cat states evolve smoothly, generating a manifold of
quantum states embedded in the projective Hilbert space
\cite{pati1995geometric,provost1980riemannian}.

The geometry of this state manifold can be characterized through the Berry
connection, and the associated Berry curvature, which determines the geometric
phase accumulated during cyclic evolutions in parameter space
\cite{berry1984quantal,zwanziger1990berry,anandan1988geometric,
chaturvedi1987berry}. The Berry curvature is defined as
\begin{equation}
\mathcal F_{ij}
=
\partial_{\lambda_i}\mathcal A_j
-
\partial_{\lambda_j}\mathcal A_i ,
\label{eq29}
\end{equation}
where
\begin{equation}
\mathcal A_i
=
i\,
\langle \psi(\lambda) |
\partial_{\lambda_i}
\psi(\lambda)
\rangle
\end{equation}
is the $i$-th component of the Berry connection. Here, the indices $i$ and $j$
denote the coordinates of the control manifold, and
$|\psi(\lambda)\rangle$ represents the normalized conditional optical cat state
defined by Eqs.~(\ref{eq:FRE_cat}) and~(\ref{eq:FPB_cat}).

The Berry curvature provides a local measure of the geometric properties of the
quantum-state manifold. For a closed trajectory $\mathcal C$ in parameter space,
the accumulated geometric phase is given by
\cite{berry1984quantal,mead1992geometric}
\begin{equation}
\gamma
=
\oint_{\mathcal C}
\mathcal A_i(\lambda)\,
d\lambda_i .
\end{equation}

The geometric phase is gauge invariant modulo $2\pi$ and depends solely on the
geometry of the closed trajectory followed by the optical state in the control
manifold. In general, the geometric phase changes continuously under smooth
deformations of the path. However, the presence of quantized phase winding or
phase accumulation that remains invariant under continuous deformations may
indicate an underlying topological structure of the quantum-state manifold.

\section{Results and Discussion}
\subsection{FRE: Displacement Loops and Berry Phase}

In this section, we investigate how the family of rotating polarization ellipses influences the geometry of the conditional optical cat state through the
HHG-induced coherent-state displacements. As discussed in Sec.~II.B, the interaction-induced displacements $\Delta\alpha_x(\theta,\phi)$ and
$\Delta\alpha_y(\theta,\phi)$ are obtained from the SFA dipole of Eqs.~(\ref{delta_alpha})--(\ref{eq7}), evaluated for the driving field of Eq.~(\ref{eq9}). These quantities
represent the corresponding shifts of the coherent-state amplitudes of the two orthogonal polarization modes.

For a fixed ellipticity, varying the control parameter $\phi$ physically rotates the polarization ellipse while preserving both its shape and the total field
amplitude, as illustrated in Fig.~\ref{fig:structured_fields}(a). Consequently, the HHG response, including the harmonic spectrum, remains unchanged under variations of the
orientation parameter. However, the decomposition of the driving field into the fixed laboratory $x$- and $y$-polarized quantized modes changes continuously
with $\phi$. As a result, the coherent-state amplitudes $\alpha_x(\phi)$ and $\alpha_y(\phi)$, defined by Eqs.~(\ref{eq14}) and~(\ref{eq15}), evolve
continuously with the orientation parameter. Evaluating the interaction-induced displacements for each value of the control parameter therefore yields the
corresponding mode displacements $\Delta\alpha_x(\phi)$ and $\Delta\alpha_y(\phi)$, which also vary continuously with $\phi$. Consequently, the conditional optical cat state evolves continuously with the control parameter, defining a closed path in the quantum-state manifold over one complete cycle of $\phi$. This cyclic evolution gives rise to the accumulation of a geometric (Berry) phase.

\begin{figure}[t!]
    \centering
    \includegraphics[width=1.0\linewidth]{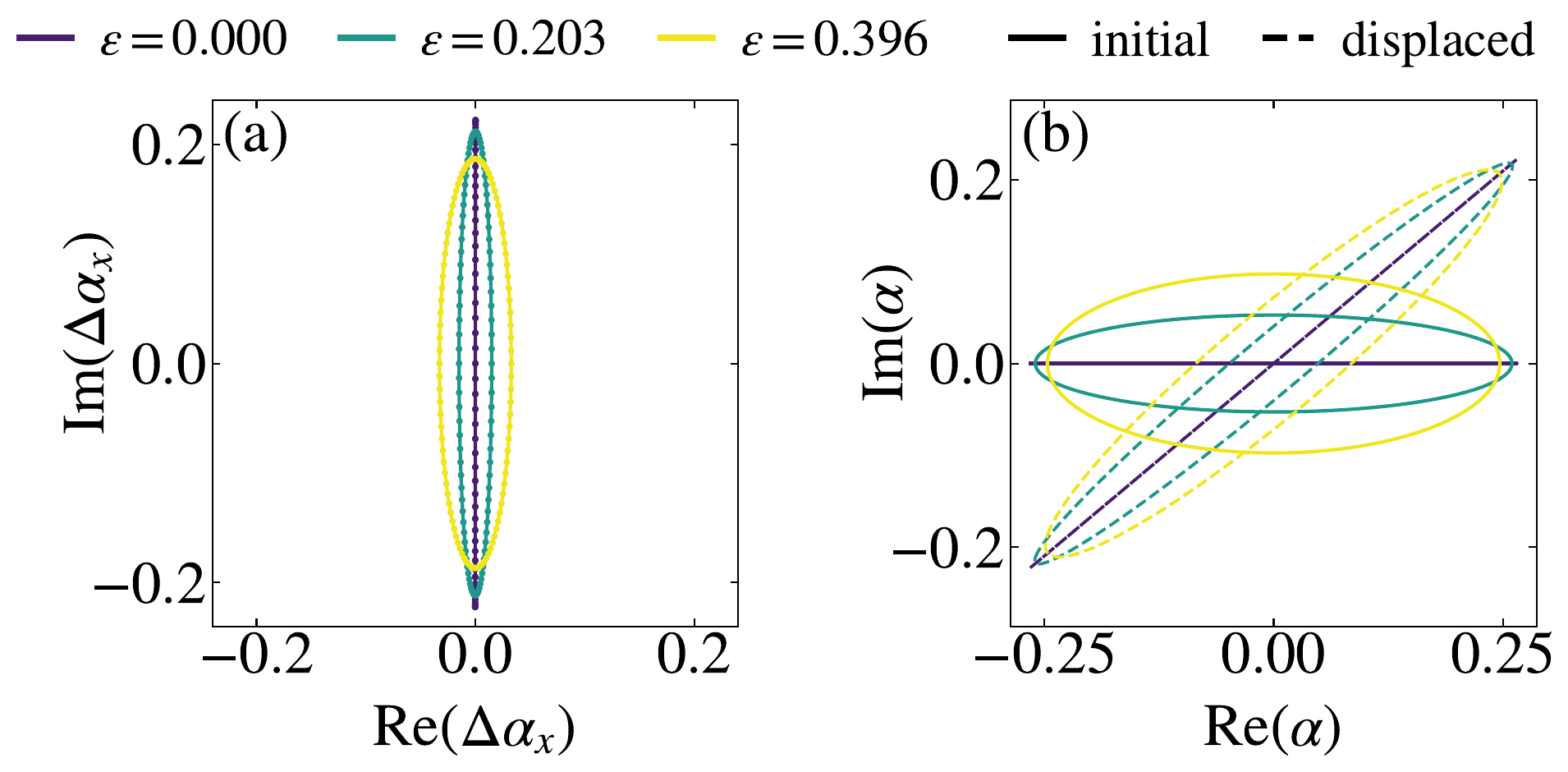}
    \caption{
(a) Complex-plane trajectories of the fundamental-mode displacement $\Delta\alpha_x(\phi)$ for different values of the ellipticity parameter $\epsilon$,
(b) Comparison between the initial coherent state $|\alpha_x(\phi)\rangle$(solid) and the displaced coherent state $|\alpha_x(\phi)+\Delta\alpha_x(\phi)\rangle$(dotted). 
Parameters used: $E_0 = 0.053$ a.u., $\omega = 0.057$ a.u., $I_p = 0.5$ a.u. and 
$n_{\mathrm{cyc}}=5$.}

\label{fig:FRE_displacement}
\end{figure}

Figure~\ref{fig:FRE_displacement}(a) shows the trajectories of the HHG-induced coherent-state displacement $\Delta\alpha_x(\phi)$ in the complex plane as the orientation parameter $\phi$ is varied over one complete cycle for selected low-to-moderate ellipticities. We restrict the
analysis to this regime because, as shown in Fig.~\ref{fig:hhg_response}(a), the HHG yield is significantly suppressed at larger ellipticities. In the linear polarization limit, the displacement remains confined to a single quadrature and, therefore, does not enclose a finite phase-space area. For finite ellipticities, both
quadratures contribute to the displacement, giving rise to closed two-dimensional trajectories in the complex plane. As the ellipticity is increased within the considered range, the geometry of these loops changes systematically, and the enclosed area increases.
Figure~\ref{fig:FRE_displacement}(b) compares the initial coherent-state trajectory, $|\alpha_x(\phi)\rangle$, with the displaced coherent-state trajectory, $|\alpha_x(\phi)+\Delta\alpha_x(\phi)\rangle$, over one complete cycle of the orientation parameter $\phi$. For finite ellipticities, the HHG-induced displacement modifies both the shape and the enclosed area of the trajectory relative to the initial coherent state. This geometric modification can be quantified by the area difference,
\[
\Delta A=A_{\mathrm{disp}}-A_{\mathrm{init}}.
\]
Since the Berry phase depends on the geometry of the closed path traced by the quantum state during cyclic evolution [Eq.~(\ref{eq29})], it is natural to examine how this HHG-induced geometric modification is reflected in the corresponding geometric phase.

\begin{figure}[t!]
    \centering
    \includegraphics[width=\linewidth]{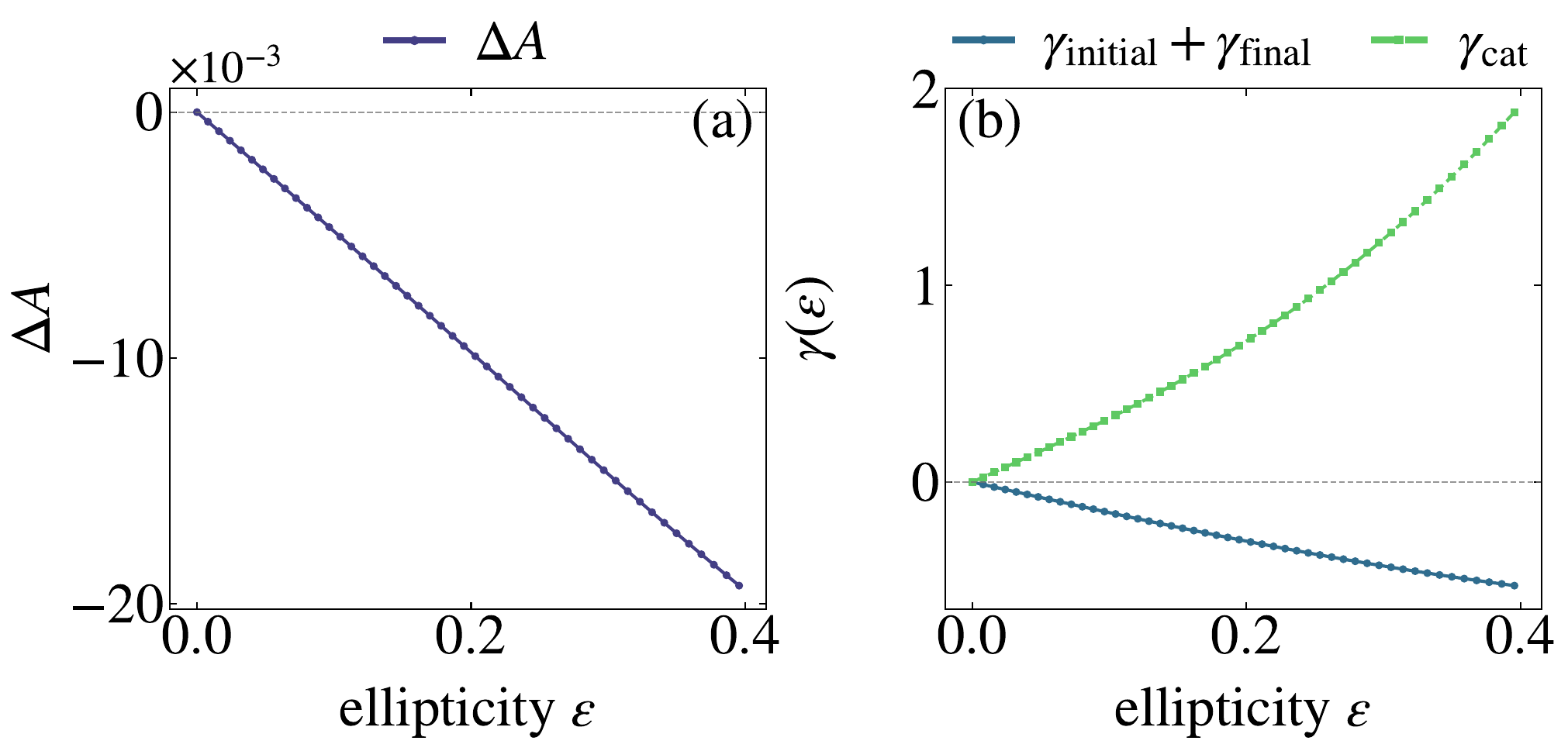}
\caption{
(a) Area difference $\Delta A=A_{\mathrm{disp}}-A_{\mathrm{init}}$ for the 
$\alpha_x$ mode, obtained from the phase-space trajectories over one complete 
cycle of the orientation parameter $\phi$, as a function of ellipticity 
$\epsilon$. 
(b) Berry phase acquired by the conditional cat state after one complete 
cycle of $\phi$ as a function of ellipticity $\epsilon$. The cat-state Berry 
phase $\gamma_{\mathrm{cat}}$ is compared with the sum of the Berry phases of 
the initial and HHG-displaced coherent states, $\gamma_{\mathrm{initial}}+
\gamma_{\mathrm{final}}$, calculated for both polarization modes 
$(\alpha_x,\alpha_y)$. The parameters are the same as those in Fig.~(\ref{fig:FRE_displacement}).
}
    \label{fig:fre_deltaA_berry_vs_ellipticity}
\end{figure}

Figure~\ref{fig:fre_deltaA_berry_vs_ellipticity}(a) shows the area difference as a function of the ellipticity. Within the range considered, $|\Delta A|$ increases monotonically, reflecting the progressively larger geometric modification of the coherent-state trajectory as the ellipticity increases. The corresponding Berry phases accumulated over one complete cycle of the orientation parameter ($\phi:0\rightarrow2\pi$) are shown in Fig.~\ref{fig:fre_deltaA_berry_vs_ellipticity}(b) for the same ellipticities.

Since the conditional cat state is constructed as a coherent superposition of the initial and HHG-displaced coherent states, it is natural to compare its Berry phase, $\gamma_{\rm cat}$, with the combined Berry phases of the constituent coherent states, $\gamma_{\rm initial}+\gamma_{\rm final}$. For an individual coherent state, the Berry phase is determined solely by the geometry of the closed trajectory traced in complex plane. This comparison therefore allows us to examine how the geometric phase of the conditional cat state is related to those of its constituent coherent-state components as the ellipticity is varied. Although both quantities exhibit the same overall dependence on ellipticity, reflecting their common geometric origin, they differ quantitatively. This difference arises from the quantum-superposition structure of the conditional cat state and the finite overlap between its coherent-state components. Consequently, the Berry phase of the conditional cat state cannot be interpreted simply as the sum of the Berry phases of its constituent coherent states.

\subsection{FPB: Mode-Resolved Response}
\begin{figure*}[t]
    \centering
    \includegraphics[width=\textwidth]{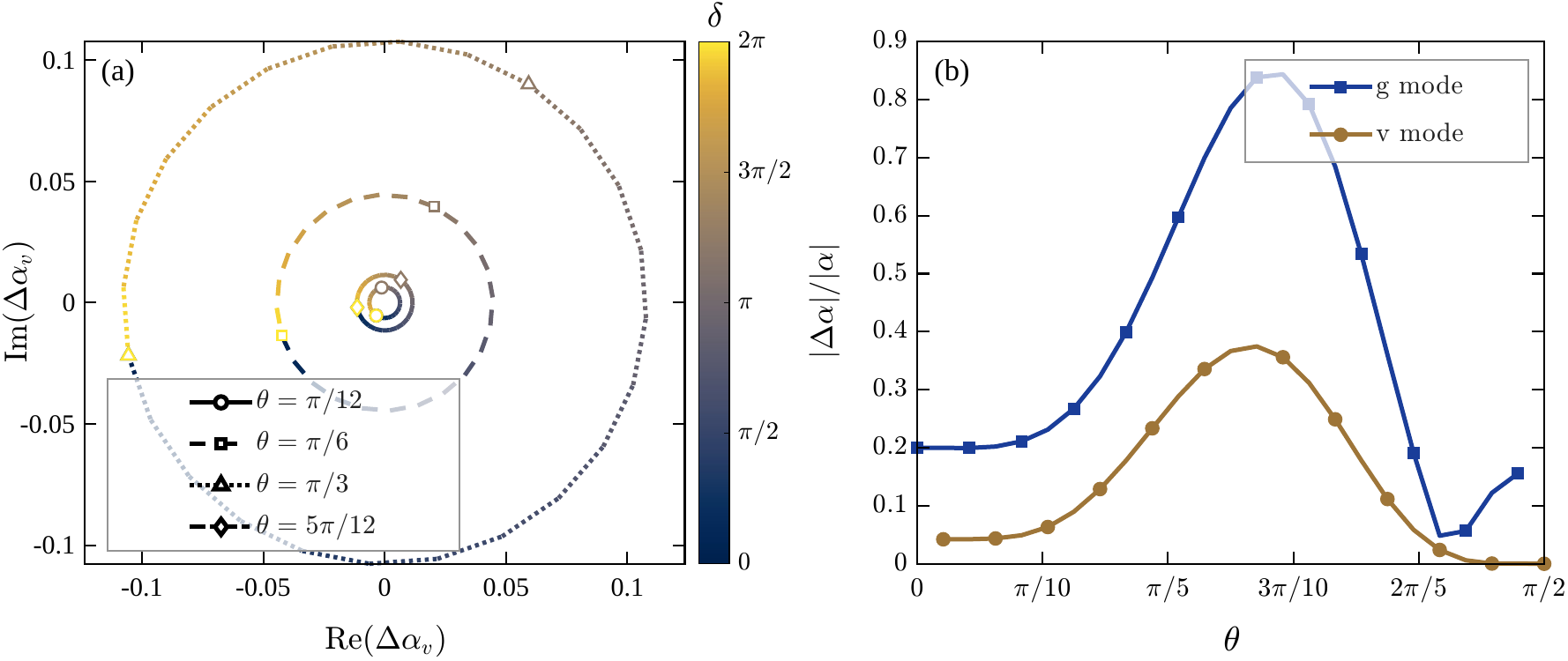}

\caption{(a) Complex-plane trajectories of the HHG-induced vortex-mode coherent-state displacement, 
$\Delta\alpha_v(\delta)$ [Eq.~(24)], for different values of the relative mode amplitude parameter 
$\theta$, as the relative phase $\delta$ is varied over one complete cycle. 
(b) Relative coherent-state displacement, $|\Delta\alpha|/|\alpha|$, for the Gaussian and vortex modes 
as a function of $\theta$, where the initial coherent-state amplitudes $|\alpha|$ are given by Eq.~(\ref{FPB_amplitude}). 
The calculations were performed using $E_0 = 0.5$ a.u., $\omega = 0.057$ a.u., 
$n_A = 1\times10^{8}~\mathrm{mm}^{-2}$,$I_p = 0.5$ a.u., 
$w_0 = 0.3125~\mathrm{mm}$ (millimeter),  $\rho = 4\omega_0$ 
and $n_{\mathrm{cyc}} = 5$.}
    \label{fig:fpb_displacement}
\end{figure*}

We now investigate the HHG-induced coherent-state displacements generated by the full Poincar\'e beam (FPB). The driving field consists of Gaussian and vortex spatial modes with opposite circular polarizations, whose coherent superposition produces the spatially varying polarization distribution shown in Fig.~\ref{fig:structured_fields}(b). Consequently, the local polarization state, and hence the driving-field ellipticity, varies across the beam profile. The relative amplitudes of the two modes are controlled by the parameter $\theta$, while the relative phase $\delta$ serves as the cyclic parameter for the geometric evolution. Figure~\ref{fig:fpb_displacement} summarizes the dependence of the mode-resolved coherent-state displacements on the relative mode amplitudes.

Figure~\ref{fig:fpb_displacement}(a) shows the complex-plane trajectories of the vortex-mode displacement, $\Delta\alpha_v(\delta)$, as the relative phase $\delta$ is varied over one complete cycle. Similar to the initial vortex coherent state (Eq.~(\ref{FPB_amplitude})), the HHG-induced displacement inherits the characteristic phase dependence on $\delta$, producing closed loops in the complex plane. The size of these loops depends on the relative amplitudes of the Gaussian and vortex modes, determined by the parameter $\theta$. In contrast, the Gaussian-mode displacement, $\Delta\alpha_g$, is independent of the relative phase $\delta$ and depends only on the relative amplitudes of the two modes through $\theta$.

\begin{figure*}[!ht]
    \centering
    \includegraphics[width=\textwidth]{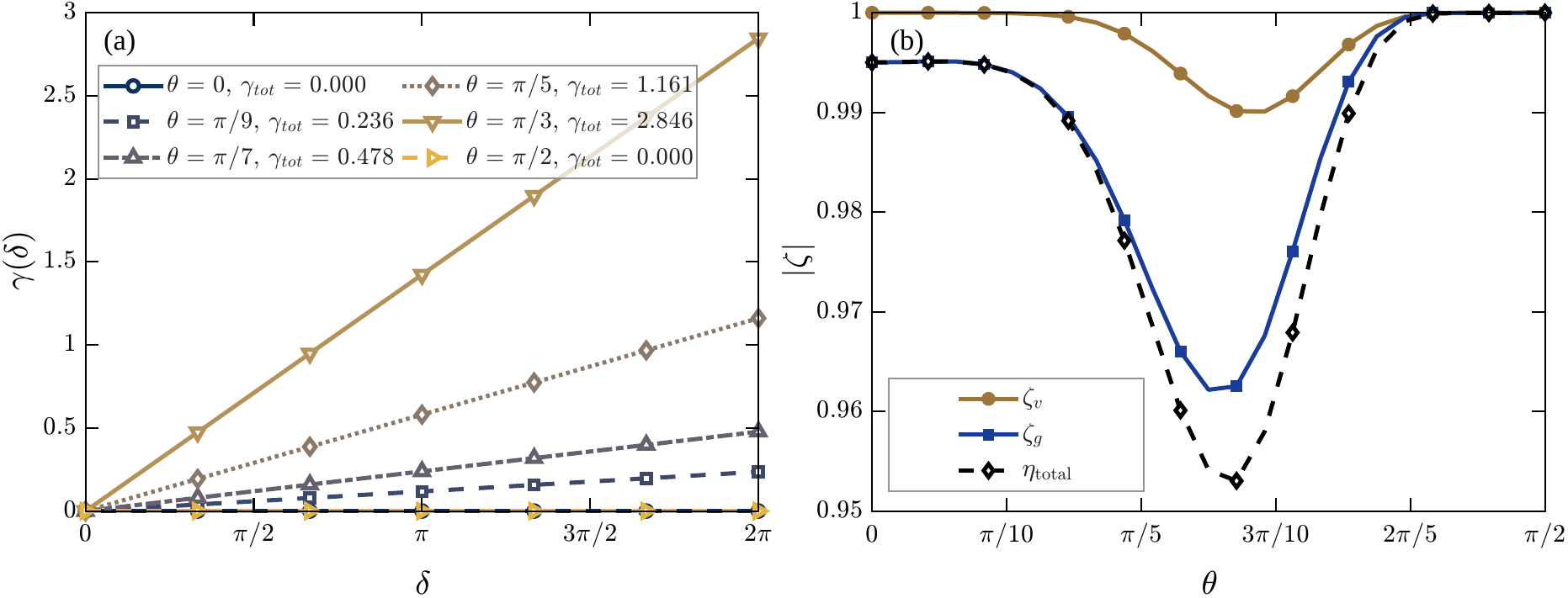}
\caption{(a) Accumulated Berry phase, $\gamma(\delta)$, of the HHG-generated optical cat state as the relative phase $\delta$ is varied over one complete cycle for different values of the relative mode amplitude parameter $\theta$. (b) Coherent-state overlap, $\zeta=\langle\alpha|\alpha-\Delta\alpha\rangle$ [Eq.~(\ref{overlap})], for the Gaussian, vortex, and total optical cat states as a function of $\theta$. The parameters are the same as those used in Fig.~\ref{fig:fpb_displacement}.}
    \label{fig:berry_phase}
\end{figure*}

Figure~\ref{fig:fpb_displacement}(b) summarizes the relative coherent-state displacement, $|\Delta\alpha|/|\alpha|$, for both the Gaussian and vortex modes as a function of $\theta$. Both modes exhibit the same qualitative behavior: the relative displacement is strongly reduced in the two limiting cases, $\theta \rightarrow 0$ and $\theta \rightarrow \pi/2$, where the Gaussian and vortex components dominate, respectively. In these limits, the driving field approaches an almost purely circular polarization, suppressing electron recollision and consequently reducing the HHG-induced coherent-state displacement. As the relative amplitudes of the two modes become comparable, the effective ellipticity of the driving field decreases, enhancing electron recollision and increasing the HHG-induced coherent-state displacement.

Finally, Fig.~\ref{fig:berry_phase}(a) shows the accumulation of the Berry phase, $\gamma(\delta)$, as the relative phase $\delta$ is varied over one complete cycle for different values of $\theta$. No Berry phase is accumulated in the limiting cases, $\theta \rightarrow 0$ and $\theta \rightarrow \pi/2$, where the vortex-mode displacement becomes negligible due to the weak HHG response. This is further confirmed by the coherent-state overlap defined in Eq.~(\ref{overlap}), shown in Fig.~\ref{fig:berry_phase}(b), where the vortex-mode overlap approaches unity as $\theta \rightarrow 0$ and $\theta \rightarrow \pi/2$, indicating that the initial and final coherent states become nearly identical. In this regime, the HHG response is too weak to produce the two distinct coherent states required for the cat-state construction, and consequently no geometric phase is accumulated. At intermediate values of $\theta$, the Gaussian and vortex components have comparable amplitudes, which reduces the effective local ellipticity of the driving field over a significant part of the transverse beam profile. This enhances electron recollision and consequently increases the HHG-induced dipole response. The resulting larger vortex-mode displacement increases the area enclosed by its trajectory in phase space, leading to a larger accumulated geometric phase. Since this displacement varies linearly with the relative phase $\delta$, the accumulated Berry phase also increases linearly with $\delta$, reaching its maximum for intermediate relative mode amplitudes.

\section{Experimental feasibility}

A key issue is whether the proposed protocol can be implemented using currently available experimental technology. To address this question, we examine its feasibility in the context of existing HHG experiments, identify the additional requirements imposed by structured driving fields and assess the principal technical challenges associated with the implementation. 

The laser parameters adopted in this work closely match those routinely used in tabletop HHG experiments. Specifically, a carrier wavelength of $800$ nm with a peak intensity of $10^{14}$ W/cm$^2$, and a pulse duration of approximately $13.3$ fs are considered. These parameters are well within the capabilities of modern chirped-pulse amplification systems and are consistent with those employed in recent structured-light HHG experiments \cite{hernandez2017extreme}. While both quantum-spectrometer-based HHG measurements \cite{lewenstein2021generation} and structured-light generation in the ultrafast realm have been independently demonstrated \cite{hernandez2017extreme,otte2018spatial,fang2025ultrafast, forbes2021structured, dorney2019controlling, de2022extreme, geneaux2016synthesis,kaarcher2023generation,dong2026generation}, the novelty of the proposed experiment lies in combining HHG-based engineer and control of nonclassical states of light with driving fields possessing a non-trivial spatial polarization topology. Their integration entails specific experimental requirements related to the synthesis, stabilization, and characterization of complex driving fields, as well as to the correlation-based detection of the conditioned quantum states. 

Unlike conventional HHG schemes reported in the literature, the present approach relies on the dynamical tuning of structured driving-field parameters, including continuously varying polarization textures, to access a multidimensional control space for tailoring the coherent-state displacement induced by the HHG process. Furthermore, stable control of the relative phase between orthogonal modes is essential, avoiding phase jitter caused by delay-line instabilities, environmental perturbations, thermal effects in the optics, and long-term mechanical drift. 

The generation of  ultrafast structured laser beams, characterized by tailored spatial phase and polarization distributions, has become a rapidly developing research field during the last decade,  leading to the emergence of several mature optical techniques that are now available for high-power applications \cite{otte2018spatial, fang2025ultrafast}.  These techniques can be broadly classified into passive and active approaches. Passive techniques, including liquid-crystal polarization optics, q-plates, phase plates and diffractive optical elements, offer robust operation and excellent power handling while requiring little active stabilization \cite{hernandez2017extreme,dorney2019controlling, de2022extreme, geneaux2016synthesis, kaarcher2023generation, schmidt2025self, turpin2017extreme, martin2025extreme}. In contrast, active approaches based on spatial light modulators (SLMs) provide dynamic and highly versatile control over the amplitude, phase and polarization of the driving field,  making them particularly attractive for experiments requiring  continuous tuning of the structured-light parameters and reconfigurable beam-shaping \cite{camper2014high}.  

Among the active approaches, the holographic multiplexing technique \cite{otte2018spatial} represents a particularly promising approach for the generation of FPBs. Unlike conventional interferometric schemes, which rely on the coherent superposition of multiple optical modes, this method employs a single phase-only SLM to directly engineer the desired beam profile. By means of spatial multiplexing, amplitude, phase and polarization can be simultaneously tailored while retaining the full spatial resolution of the modulator. In addition, the use of a single optical component further enhances stability and mitigates the impact of long-term phase fluctuations, thereby providing a robust highly reproducible platform for the synthesis of the complex polarization distributions.  

A fundamental requirement for the generation of structured polarization states is the ability to independently control the relative amplitude and phase of two orthogonal field components. In ultrafast optics, phase-locked pulse replicas in the visible and infrared spectral regions are commonly generated using amplitude-division interferometers, such as Michelson or Mach-Zehnder configurations. Similar architectures are widely employed in attosecond pump-probe experiments \cite{chini2009delay, csizmadia2025active}, where they enable temporal resolution on sub-femtosecond timescales. They have also been extensively used in high-harmonic generation to synthesize tailored driving waveforms through the coherent combination of multiple frequency components \cite{burger2017compact, raab2024highly}. However, because interferometric measurements are highly sensitive to optical path-length fluctuations, even displacements corresponding to a small fraction of the optical wavelength can significantly degrade the interference contrast and compromise phase stability. As a result, active stabilization of the interferometer is generally needed.

Common-path optical architectures offer an attractive alternative. By guiding the pulse replicas along nearly identical trajectories, these schemes intrinsically suppress differential optical path fluctuations while retaining precise control of the relative delay with sub-cycle resolution \cite{jansen2016spatially,ardini2024generation, kuzkova2026attosecond}. As a result, they can achieve excellent passive phase stability without the need for complex active stabilization systems. Furthermore, their intrinsic robustness makes them particularly well suited for experiments requiring long acquisition times, such as correlation-based quantum measurements, where reproducibility and long-term stability are essential. 

A crucial aspect of structured-light-driven HHG concerns the preservation of the polarization topology throughout beam propagation and focusing. Polarization distortions may arise from the differential response of focusing and beam-folding optics, which can introduce relative phase shifts between the orthogonal s- and p-polarized field components. In addition, under tight-focusing conditions, the emergence of longitudinal electric-field components can significantly modify the local vectorial structure of the beam \cite{garcia2024topological, watzel2020multipolar}. Both effects may alter the intended structured field and consequently influence the local strong-field dynamics. Although polarization-preserving focusing is routinely achieved in vector-beam optics, its extension to HHG requires careful ptimization of the optical layout and accurate alignment and calibration procedures \cite{hernandez2017extreme, fang2025ultrafast, forbes2021structured}. Reflective focusing optics, such as off-axis parabolic mirrors commonly employed in attosecond beamlines , provide well suited solution, combining achromatic focusing with excellent preservation of the spatial polarization distribution. 

A further requirement is the complete characterization of the structured driving field prior to the HHG interaction. Since the proposed method relies on controlled trajectories within a multidimensional parameter space, accurate diagnostics are essential to establish a direct correspondence between the experimentally generated optical field and the driving field eassumed in the stheretical model. Conventional HHG experiments generally require characterization of the pulse duration, energy, spatial profile and wavefront. In the present scheme, however, the spatially varying polarization state constitues an additional degree of freedom that directly governs the trajectory of the coherent-state displacement. The experiment should, therefore, include spatially resolved polarimetry capable of reconstructing the Stokes parameters across the beam profile, complemented by wavefront sensing and spatial interferometry to verify the amplitude and phase distributions of the generated structured beam \cite{otte2018spatial, fang2025ultrafast}.  

Concerning the quantum measurement stage, it has been previously reported for a single-color, linearly polarized driving IR field \cite{lewenstein2021generation}. The protocol developed for implementing conditional measurements is a direct consequence of the correlation and energy conservation between the IR photon losses and the emitted harmonic photons. Developing protocols for implementing conditional measurements when structured driving IR light fields are used is more challenging and lies beyond the scope of the present work. This is because addressing this problem requires a complete understanding of the correlations between the energy and the degrees of freedom of the structured light field and the emitted harmonics. In other words, for each type of structured light field, a new protocol must be developed. 

\section{Conclusion and Outlook} 
The present results show that structured driving fields provide a powerful mechanism for controlling the geometric properties of HHG-generated conditional cat states. By tuning the polarization and spatial structure of the driving field, one can modify the phase-space trajectories, enclosed areas, and the associated geometric phases of the generated states. Since geometric phases are physically observable through interference effects, this opens the possibility of manipulating the phase structure and coherence properties of conditional cat states through external control parameters. More generally, our work highlights HHG in gas phase as a platform for engineering and exploring nontrivial geometric features of macroscopic quantum superposition states.

It would be interesting to extend the present results to structured fields involving two frequencies. Such fields may themselves exhibit nontrivial topological properties, such as fractional order knots in the polarization space \cite{pisanty2019,pisanty2019conservation}, and others \cite{kong2017controlling,paufler2019high}. Such fields suggest a possibility of generating truly topological Schrödinger cat states via post-selection. The technical problem here is that such structured fields have a typically non-uniform spatial intensity distribution. In effect, one is obliged to perform a spatial average, as for example presented in Eqs.~(\ref{23})-(\ref{25}), or to focus on harmonics generated locally, from a given region. Both approaches might be detrimental to the possibility of observing topological cat states.

\begin{acknowledgments}
We thank Jens Biegert, Phil Bucksbaum, Carla Faria and Lidija Petrovic, for discussions. ICFO-QOT group acknowledges support from:
European Research Council AdG NOQIA; MCIN/AEI (PGC2018-0910.13039/501100011033,  CEX2019-000910-S/10.13039/501100011033, Plan National FIDEUA PID2019-106901GB-I00, Plan National STAMEENA PID2022-139099NB, I00, project funded by MCIN / AEI / 10.13039 / 501100011033 and by the “European Union NextGenerationEU/PRTR" (PRTR-C17.I1), FPI); QUANTERA DYNAMITE PCI2022-132919, QuantERA II Programme co-funded by European Union’s Horizon 2020 program under Grant Agreement No 101017733; Ministry for Digital Transformation and of Civil Service of the Spanish Government through the QUANTUM ENIA project call - Quantum Spain project, and by the European Union through the Recovery, Transformation and Resilience Plan - NextGenerationEU within the framework of the Digital Spain 2026 Agenda;
Fundació Cellex; Fundació Mir-Puig; Generalitat de Catalunya (European Social Fund FEDER and CERCA program; Barcelona Supercomputing Center MareNostrum (FI-2023-3-0024); Funded by the European Union. Views and opinions expressed are however those of the author(s) only and do not necessarily reflect those of the European Union, European Commission, European ClimateInfrastructure and Environment Executive Agency (CINEA), or any other granting authority.  Neither the European Union nor any granting authority can be held responsible for them (HORIZON-CL4-2022-QUANTUM-02-SGA  PASQuanS2.1, 101113690, EU Horizon 2020 FET-OPEN OPTOlogic, Grant No 899794, QU-ATTO, 101168628),  EU Horizon Europe Program (This project has received funding from the European Union’s Horizon Europe research and innovation program under grant agreement No 101080086 NeQSTGrant Agreement 101080086 — NeQST); 
ICFO Internal “QuantumGaudi” project.

P.T. acknowledges the European Union’s HORIZONMSCA-2023-DN-01 project QU-ATTO under the Marie Skłodowska-Curie grant agreement No 101168628 and ELI–ALPS. ELI–ALPS is supported by the EU and cofinanced by the European Regional Development Fund (GINOP No. 2.3.6-15-2015-00001). 

M. F. C. acknowledges support by the Quantum Science and Technology-National Science and Technology Major Project (Grant No. 2025ZD0301000),  the National Key Research and Development Program of China (Grant No. 2023YFA1407100), the Guangdong Province Science and Technology Major Project (Future functional materials under extreme conditions - 2021B0301030005) and the National Natural Science Foundation of China (Grant No. 12574092).
E.P.\ acknowledges Royal Society funding under URF\textbackslash{}\allowbreak{}R1\textbackslash{}\allowbreak{}211390.

\end{acknowledgments}


\section*{Data Availability Statement}

The data that support the findings of this study are available from the
corresponding author upon reasonable request.


\bibliographystyle{apsrev4-2}
\bibliography{references}

@article{GTB22,
  title   = {\href{https://www.nature.com/articles/s41567-023-02127-y}
  {High-harmonic generation driven by quantum light}},
  author  = {Gorlach, Alexey and Tzur, Matan Even and Birk, Michael and Kr{\"u}ger, Michael and Rivera, Nicholas and Cohen, Oren and Kaminer, Ido},
  journal = {Nat. Phys.},
  volume  = {19},
  number  = {11},
  pages   = {1689--1696},
  year    = {2023}
}

@article{corkum1993plasma,
  author = {Paul B. Corkum},
  title = {\href{https://doi.org/10.1103/PhysRevLett.71.1994}{Plasma perspective on strong field multiphoton ionization}},
  journal = {Phys. Rev. Lett.},
  volume = {71},
  number = {13},
  pages = {1994--1997},
  year = {1993}
}

@article{gonoskov2016quantum,
  author = {Igor A. Gonoskov and Nikolaos Tsatrafyllis and Ioannis K. Kominis and Paraskevas Tzallas},
  title = {\href{https://doi.org/10.1038/srep32821}{Quantum optical signatures in strong-field laser physics: Infrared photon counting in high-order-harmonic generation}},
  journal = {Sci. Rep.},
  volume = {6},
  number = {1},
  pages = {32821},
  year = {2016}
}

@article{gorlach2020quantum,
  author = {Alexey Gorlach and Ofer Neufeld and Nicholas Rivera and Oren Cohen and Ido Kaminer},
  title = {\href{https://doi.org/10.1038/s41467-020-18343-6}{The quantum-optical nature of high harmonic generation}},
  journal = {Nat. Commun.},
  volume = {11},
  number = {1},
  pages = {4598},
  year = {2020}
}

@article{lamprou2025nonlinear,
  author = {Theocharis Lamprou and Javier Rivera-Dean and Philipp Stammer and Maciej Lewenstein and Paraskevas Tzallas},
  title = {\href{https://doi.org/10.1103/PhysRevLett.134.013601}{Nonlinear optics using intense optical coherent state superpositions}},
  journal = {Phys. Rev. Lett.},
  volume = {134},
  number = {1},
  pages = {013601},
  year = {2025}
}

@article{stammer2022theory,
  author = {Philipp Stammer},
  title = {\href{https://doi.org/10.1103/PhysRevA.106.L050402}{Theory of entanglement and measurement in high-order harmonic generation}},
  journal = {Phys. Rev. A},
  volume = {106},
  number = {5},
  pages = {L050402},
  year = {2022}
}

@article{stammer2024metrological,
  author = {Philipp Stammer and Tom\'as Fern\'andez Martos and Maciej Lewenstein and Grzegorz Rajchel-Mieldzio\'c},
  title = {\href{https://arxiv.org/abs/2311.01371}{Metrological robustness of high photon number optical cat states}},
  journal = {Quantum Sci. Technol.},
  volume = {9},
  number = {4},
  pages = {045047},
  year = {2024}
}

@article{lange2024electron,
  author = {Christian Saugbjerg Lange and Thomas Hansen and Lars Bojer Madsen},
  title = {\href{https://doi.org/10.1103/PhysRevA.109.033110}{Electron-correlation-induced nonclassicality of light from high-order harmonic generation}},
  journal = {Phys. Rev. A},
  volume = {109},
  number = {3},
  pages = {033110},
  year = {2024}
}

@article{lange2025hierarchy,
  author = {Christian Saugbjerg Lange and Lars Bojer Madsen},
  title = {\href{https://doi.org/10.1103/PhysRevA.111.013113}{Hierarchy of approximations for describing quantum light from high-harmonic generation: A Fermi-Hubbard-model study}},
  journal = {Phys. Rev. A},
  volume = {111},
  number = {1},
  pages = {013113},
  year = {2025}
}

@article{stammer2024entanglement,
  author = {Philipp Stammer and Javier Rivera-Dean and Andrew S. Maxwell and Theocharis Lamprou and Javier Arg\"uello-Luengo and Paraskevas Tzallas and Marcelo F. Ciappina and Maciej Lewenstein},
  title = {\href{https://doi.org/10.1103/PhysRevLett.132.143603}{Entanglement and squeezing of the optical field modes in high harmonic generation}},
  journal = {Phys. Rev. Lett.},
  volume = {132},
  number = {14},
  pages = {143603},
  year = {2024}
}

@article{rivera2024nonclassical,
  author = {Javier Rivera-Dean and Philipp Stammer and Alexander S. Maxwell and Theocharis Lamprou and Alejandro F. Ordó\~nez and Emilio Pisanty and Paraskevas Tzallas},
  title = {\href{https://doi.org/10.1103/PhysRevB.109.035203}{Nonclassical states of light after high-harmonic generation in semiconductors: A Bloch-based perspective}},
  journal = {Phys. Rev. B},
  volume = {109},
  number = {3},
  pages = {035203},
  year = {2024}
}

@article{stammer2023quantum,
  author = {Philipp Stammer and Javier Rivera-Dean and Andrew S. Maxwell and Theocharis Lamprou and Andr\'es Ord\'o\~nez and Marcelo F. Ciappina and Paraskevas Tzallas and Maciej Lewenstein},
  title = {\href{https://doi.org/10.1103/PRXQuantum.4.010201}{Quantum electrodynamics of intense laser-matter interactions: A tool for quantum state engineering}},
  journal = {PRX Quantum},
  volume = {4},
  number = {1},
  pages = {010201},
  year = {2023}
}

@article{stammer2025colloquium,
  author = {Philipp Stammer and Javier Rivera-Dean and Paraskevas Tzallas and Marcelo F. Ciappina and Maciej Lewenstein},
  title = {\href{https://arxiv.org/abs/2510.19045}{Colloquium: Quantum optics of intense light--matter interaction}},
  journal = {arXiv},
  pages = {arXiv:2510.19045},
  year = {2025}
}

@article{yi2025generation,
  author = {Sili Yi and Nikolai D. Klimkin and Graham Gardiner Brown and Olga Smirnova and Serguei Patchkovskii and Ihar Babushkin and Misha Ivanov},
  title = {\href{https://doi.org/10.1103/PhysRevX.15.011023}{Generation of massively entangled bright states of light during harmonic generation in resonant media}},
  journal = {Phys. Rev. X},
  volume = {15},
  number = {1},
  pages = {011023},
  year = {2025}
}

@article{stammer2025theory,
  author = {Philipp Stammer and Javier Rivera-Dean and Maciej Lewenstein},
  title = {\href{https://arxiv.org/abs/2504.13287}{Theory of quantum optics and optical coherence in high harmonic generation}},
  journal = {arXiv},
  pages = {arXiv:2504.13287},
  year = {2025}
}

@article{stammer2026photon,
  author  = {Stammer, Philipp and Rivera-Dean, Javier and Lewenstein, Maciej},
  title   = {\href{https://arxiv.org/abs/2606.17620}{Photon Antibunching in High-Harmonic Generation}},
  journal = {arXiv:2606.17620},
  year    = {2026}
}

@article{stammer2026attosecond,
  title={\href{https://arxiv.org/abs/2607.06395}{Attosecond metrology of bright quantum light}},
  author={Stammer, P and Rivera-Dean, J and Pisanty, E and Lewenstein, M},
  journal={arXiv:2607.06395},
  year={2026}
}

@article{rivera2024quantum,
  author = {Javier Rivera-Dean and Theocharis Lamprou and Emilio Pisanty and Marcelo F. Ciappina and Paraskevas Tzallas and Maciej Lewenstein and Philipp Stammer},
  title = {\href{https://doi.org/10.1103/PhysRevA.112.013110}{Quantum state engineering of light using intensity measurements and post-selection}},
  journal = {Phys. Rev. A},
  volume = {112},
  pages = {013110},
  year = {2025}
}

@article{Gao2020,
  author = {Sijia Gao and J{\"o}rg B. G{\"o}tte and Fiona C. Speirits and Francesco Castellucci and Sonja Franke-Arnold and Stephen M. Barnett},
  title = {\href{https://doi.org/10.1103/PhysRevA.102.053513}{Paraxial Skyrmionic Beams}},
  journal = {Phys. Rev. A},
  volume = {102},
  number = {5},
  pages = {053513},
  year = {2020},
  publisher = {APS}
}

@article{kong2017controlling,
  author = {Fanqi Kong and Chunmei Zhang and Fr\'ed\'eric Bouchard and Zhengyan Li and Graham G. Brown and Dong Hyuk Ko and T. J. Hammond and Ladan Arissian and Robert W. Boyd and Ebrahim Karimi and others},
  title = {\href{https://doi.org/10.1038/ncomms14970}{Controlling the orbital angular momentum of high harmonic vortices}},
  journal = {Nat. Commun.},
  volume = {8},
  number = {1},
  pages = {14970},
  year = {2017}
}

@article{paufler2019high,
  author = {Willi Paufler and Birger Böning and Stephan Fritzsche},
  title = {\href{https://iopscience.iop.org/article/10.1088/2040-8986/ab31c3?utm_source=researchgate.net&utm_medium=article}{High harmonic generation with Laguerre--Gaussian beams}},
  journal = {J. Opt.},
  volume = {21},
  number = {9},
  pages = {094001},
  year = {2019}
}

@article{smirnova2014multielectron,
  author = {Olga Smirnova and Misha Ivanov},
  title = {\href{https://arxiv.org/abs/1304.2413}{Multielectron high harmonic generation: Simple man on a complex plane}},
  journal = {Attosecond and XUV Physics: Ultrafast Dynamics and Spectroscopy},
  pages = {201--256},
  year = {2014}
}

@article{bhandari1997polarization,
  author = {Rajendra Bhandari},
  title = {\href{https://www.sciencedirect.com/science/article/abs/pii/S0370157396000294?via%3Dihub}{Polarization of light and topological phases}},
  journal = {Phys. Rep.},
  volume = {281},
  number = {1},
  pages = {1--64},
  year = {1997}
}

@article{yao2011orbital,
  author = {Alison M. Yao and Miles J. Padgett},
  title = {\href{https://doi.org/10.1364/AOP.3.000161}{Orbital angular momentum: Origins, behavior and applications}},
  journal = {Adv. Opt. Photon.},
  volume = {3},
  number = {2},
  pages = {161--204},
  year = {2011}
}

@article{pati1995geometric,
  author = {Arun Kumar Pati},
  title = {\href{https://doi.org/10.1103/PhysRevA.52.2576}{Geometric aspects of noncyclic quantum evolutions}},
  journal = {Phys. Rev. A},
  volume = {52},
  number = {4},
  pages = {2576--2584},
  year = {1995}
}

@article{provost1980riemannian,
  author = {Jean-Pierre Provost and G. Vall\'ee},
  title = {\href{https://doi.org/10.1007/BF02193559}{Riemannian structure on manifolds of quantum states}},
  journal = {Commun. Math. Phys.},
  volume = {76},
  number = {3},
  pages = {289--301},
  year = {1980}
}

@article{berry1984quantal,
  author = {Michael Victor Berry},
  title = {\href{https://doi.org/10.1098/rspa.1984.0023}{Quantal phase factors accompanying adiabatic changes}},
  journal = {Proc. R. Soc. Lond. A},
  volume = {392},
  number = {1802},
  pages = {45--57},
  year = {1984}
}

@article{lange2026high,
  title   = {\href{https://doi.org/10.1103/3n7b-m3vt}
  {High-Order Harmonic Generation with Beyond-Semiclassical Emitter Dynamics: A Strong-Field Quantum-Optical Heisenberg-Picture Approach}},
  author  = {Lange, Christian Saugbjerg and Lassen, Ella Elisabeth and Gothelf, Rasmus Vesterager and Madsen, Lars Bojer},
  journal = {Phys. Rev. A},
  volume  = {113},
  number  = {5},
  pages   = {053115},
  year    = {2026},
  publisher = {American Physical Society}
}

@article{gan2026tailoring,
  title   = {\href{https://arxiv.org/abs/2607.21248}
  {Tailoring Optical Schr{\"o}dinger Cat States via Orientation-Dependent High-Harmonic Generation in $\mathrm{H}_2^+$}},
  author  = {Gan, Ziyang and Jiang, Wei-Chao and Chen, Ahai and Jiang, Yuhai},
  journal = {arXiv preprint arXiv:2607.21248},
  year    = {2026}
}

@article{bai2025dynamical,
  author={Ya Bai and Hanqing Xu and Ying Ma and Jingyuan Niu and Yang Jiang and Wenyang Zheng and Shuo Wang and Candong Liu and Peng Liu and Ruxin Li},
  title={\href{https://doi.org/10.21203/rs.3.rs-6566868/v1}{Dynamical Chiral High-Harmonic Generation via Subcycle Symmetry Engineering}},
  journal={Research Square Preprint},
  year={2025}
}

@article{cisowski2026geometric,
  author={Claire Cisowski},
  title={\href{https://doi.org/10.1103/gf21-tgcm}{Geometric Representation of Higher-Order Optical Modes}},
  journal={Phys. Rev. A},
  volume={113}, 
  number={6}, 
  pages={063530},
  year={2026}, publisher={American Physical Society}
}

@article{anandan1988geometric,
  author = {J. Anandan and Yakir Aharonov},
  title = {\href{https://doi.org/10.1103/PhysRevD.38.1863}{Geometric quantum phase and angles}},
  journal = {Phys. Rev. D},
  volume = {38},
  number = {6},
  pages = {1863--1870},
  year = {1988}
}

@article{zwanziger1990berry,
  author = {Josef W. Zwanziger and Marianne Koenig and Alex Pines},
  title = {\href{https://doi.org/10.1146/annurev.pc.41.100190.003125}{Berry's phase}},
  journal = {Annu. Rev. Phys. Chem.},
  volume = {41},
  pages = {601--646},
  year = {1990}
}

@article{mead1992geometric,
  author = {C. Alden Mead},
  title = {\href{https://doi.org/10.1103/RevModPhys.64.51}{The geometric phase in molecular systems}},
  journal = {Rev. Mod. Phys.},
  volume = {64},
  number = {1},
  pages = {51--85},
  year = {1992}
}

@article{USVortex,
  author={Camilo Granados and Bikash K. Das and Christian Heide and Shambhu Ghimire and Marcelo F. Ciappina},
  title={\href{https://spj.science.org/doi/abs/10.34133/ultrafastscience.0100}{Toward Attosecond Vortices in Semiconductor Materials}},
  journal={Ultrafast Sci.},
  volume={5}, pages={0100},
  year={2025}
}

@article{schrodinger1935gegenwartige,
  author = {Erwin Schr\"odinger},
  title = {\href{https://doi.org/10.1007/BF01491891}{Die gegenw\"artige Situation in der Quantenmechanik}},
  journal = {Naturwissenschaften},
  volume = {23},
  number = {50},
  pages = {844--849},
  year = {1935}
}

@article{yurke1986generating,
  author = {Bernard Yurke and David Stoler},
  title = {\href{https://doi.org/10.1103/PhysRevLett.57.13}{Generating quantum mechanical superpositions of macroscopically distinguishable states via amplitude dispersion}},
  journal = {Phys. Rev. Lett.},
  volume = {57},
  number = {1},
  pages = {13--16},
  year = {1986}
}

@article{forbes2019structured,
  author = {Andrew Forbes},
  title = {\href{https://doi.org/10.1002/lpor.201900140}{Structured light from lasers}},
  journal = {Laser Photonics Rev.},
  volume = {13},
  number = {11},
  pages = {1900140},
  year = {2019}
}

@incollection{allen1999iv,
  author = {Les Allen and Miles J. Padgett and M. Babiker},
  title = {\href{https://doi.org/10.1016/S0079-6638(08)70391-3}{The orbital angular momentum of light}},
  booktitle = {Prog. Opt.},
  publisher={Elsevier},
  volume = {39},
  pages = {291--372},
  year = {1999}
}

@article{Beckley2010,
  author = {Amber M. Beckley and Thomas G. Brown and Miguel A. Alonso},
  title = {\href{https://doi.org/10.1364/OE.18.010777}{Full Poincar{\'e} Beams}},
  journal = {Opt. Express},
  volume = {18},
  number = {10},
  pages = {10777--10785},
  year = {2010},
  month = {May}
}

@article{milione2011higher,
  author = {Giovanni Milione and H. I. Sztul and D. A. Nolan and Robert R. Alfano},
  title = {\href{https://doi.org/10.1103/PhysRevLett.107.053601}{Higher-Order Poincar{\'e} Sphere, Stokes Parameters, and the Angular Momentum of Light}},
  journal = {Phys. Rev. Lett.},
  volume = {107},
  number = {5},
  pages = {053601},
  year = {2011},
  publisher = {APS}
}

@article{chaturvedi1987berry,
  author = {S. Chaturvedi and N. Mukunda and R. Simon},
  title = {\href{https://doi.org/10.1103/PhysRevLett.58.555}{Berry's phase for coherent states}},
  journal = {Phys. Rev. Lett.},
  volume = {58},
  number = {6},
  pages = {555--558},
  year = {1987}
}

@article{pisanty2019,
  author = {Emilio Pisanty and Gerard Jim\'enez and Ver\'onica Vicu\~na-Hern\'andez and Antonio Pic\'on and Alessio Celi and Juan P. Torres and Maciej Lewenstein},
  title = {\href{https://doi.org/10.1038/s41566-019-0451-2}{Knotting fractional-order knots with the polarization state of light}},
  journal = {Nat. Photonics},
  volume = {13},
  pages = {569--574},
  year = {2019}
}

@article{pisanty2019conservation,
  author = {Emilio Pisanty and Laura Rego and Julio San Rom\'an and Antonio Pic\'on and Kevin M. Dorney and Henry C. Kapteyn and Margaret M. Murnane and Luis Plaja and Maciej Lewenstein and Carlos Hern\'andez-Garc\'ia},
  title = {\href{https://doi.org/10.1103/PhysRevLett.122.203201}{Conservation of torus-knot angular momentum in high-order harmonic generation}},
  journal = {Phys. Rev. Lett.},
  volume = {122},
  number = {20},
  pages = {203201},
  year = {2019}
}

@article{glauber1963coherent,
  title   = {\href{https://journals.aps.org/pr/abstract/10.1103/PhysRev.131.2766}
  {Coherent and incoherent states of the radiation field}},
  author  = {Glauber, Roy J.},
  journal = {Phys. Rev.},
  volume  = {131},
  number  = {6},
  pages   = {2766--2788},
  year    = {1963},
  publisher = {American Physical Society}
}

@book{Walls&Milburn,
  title={Quantum Optics},
  author={Walls, D. F. and Milburn, G. J.},
  year={2010},
  publisher={Springer Berlin, Heidelberg}
}

@article{RLP22,
  title   = {\href{https://journals.aps.org/pra/abstract/10.1103/PhysRevA.105.033714}
  {Strong laser fields and their power to generate controllable high-photon-number coherent-state superpositions}},
  author  = {Rivera-Dean, J. and Lamprou, Th. and Pisanty, E. and Stammer, P. and Ord{\'o}{\~n}ez, A. F. and Maxwell, A. S. and Ciappina, M. F. and Lewenstein, M. and Tzallas, P.},
  journal = {Phys. Rev. A},
  volume  = {105},
  number  = {3},
  pages   = {033714},
  year    = {2022},
  publisher = {American Physical Society}
}

@article{SRL22,
  title   = {\href{https://journals.aps.org/prl/abstract/10.1103/PhysRevLett.128.123603}
  {High Photon Number Entangled States and Coherent State Superposition from the Extreme Ultraviolet to the Far Infrared}},
  author  = {Stammer, Philipp and Rivera-Dean, Javier and Lamprou, Th. and Pisanty, Emilio and Ciappina, Marcelo F. and Tzallas, Paraskevas and Lewenstein, Maciej},
  journal = {Physical Review Letters},
  volume  = {128},
  number  = {12},
  pages   = {123603},
  year    = {2022},
  publisher = {American Physical Society}
}

@article{SRC25,
  title={\href{https://arxiv.org/abs/2508.09048}{Weak measurement in strong laser field physics}},
  author={Stammer, Philipp and Rivera-Dean, Javier and Ciappina, Marcelo F and Lewenstein, Maciej},
  journal={arXiv:2508.09048},
  year={2025}
}

@incollection{kulander_dynamics_1993,
  author = {K. C. Kulander and K. J. Schafer and J. L. Krause},
  title = {Dynamics of short-pulse excitation, ionization and harmonic conversion},
  booktitle = {\href{https://www.springer.com/gp/book/9780306445873}{Super-Intense Laser Atom Physics}},
  editor = {B. Piraux and A. L'Huillier and K. Rz\k{a}\.zewski},
  series = {NATO Advanced Studies Institute Series B: Physics},
  publisher = {Plenum},
  address = {New York},
  volume = {316},
  pages = {95--110},
  year = {1993},
}

@article{lewenstein1994theory,
  title={\href{https://journals.aps.org/pra/abstract/10.1103/PhysRevA.49.2117}
  {Theory of high-harmonic generation by low-frequency laser fields}},
  author={Lewenstein, Maciej and Balcou, Ph. and Ivanov, M. Yu. and L'Huillier, Anne and Corkum, Paul B.},
  journal={Phys. Rev. A},
  volume={49},
  number={3},
  pages={2117--2132},
  year={1994},
  publisher={American Physical Society}
}

@preamble{ "\providecommand{\noopsort}[1]{} " }

@article{Ido-Nirit-new,
  title = {\href{https://arxiv.org/abs/2511.18362}{Attosecond-resolved quantum fluctuations of light and matter}},
  author = {Tzur, Matan Even and Mor, Chen and Yaffe, Noa and Birk, Michael and Rasputnyi, Andrei and Kneller, Omer and Nisim, Ido and Kaminer, Ido and Chekhova, Maria and Krueger, Michael and others},
    journal={arXiv preprint arXiv:2511.18362},
  year = {2025},
 
}

@article{Javier-ATI-BSL,
  author={Javier Rivera-Dean and Philipp Stammer and Carla Figueira de Morisson Faria and Maciej Lewenstein},
  title={\href{https://journals.aps.org/pra/abstract/10.1103/hb3n-h2zy}{Microscopic analysis of above-threshold ionization driven by squeezed light}},
  journal={Phys. Rev. A},
  volume={112}, pages={063101},
  year={2025}
}

@article{paris-BSL-prop,
  author={Javier Rivera-Dean and D. Kanti and Philipp Stammer and N. Tsatrafyllis and Maciej Lewenstein and Paraskevas Tzallas},
  title={\href{https://arxiv.org/abs/2509.19608}{Propagation of intense squeezed vacuum light in non-linear media}},
  journal={arXiv preprint arXiv:2509.19608},
  year={2025}
}

@article{Lidija-1-mode,
  author={Lidija Petrovic and Philipp Stammer and Maciej Lewenstein and Javier Rivera-Dean},
  title={\href{https://arxiv.org/abs/2601.01611}{Generation of circular polarized high-order harmonics from single color quantum light}},
  journal={Phys. Rev. A},
  volume={114},
  number={1},
  pages={013127},
  year={2026},
  publisher={APS}
}

@article{Gauss-Technion,
  author = {J. Rivera-Dean and M. Even-Tzur and M. F. Ciappina and C. Granados and O. Cohen and P. Stammer},
  title = {\href{https://arxiv.org/abs/2603.24377}{Emergence of Gaussian entanglement and non-Gaussianity in high-harmonic generation driven by bright squeezed light}},
  journal = {arXiv:2606.22714},
  year = {2026},
  arxiv = {arXiv:2606.22714}
}

@article{Stammer_EnergyConservation2024,
  title   = {\href{https://arxiv.org/abs/2410.15503}
  {Energy conservation in quantum optical high harmonic generation}},
  author  = {Stammer, P.},
  journal = {arXiv:2410.15503},
  year    = {2024}
}

@article{TCS24,
  author={D. Theidel and V. Cotte and R. Sondenheimer and V. Shiriaeva and M. Froidevaux and V. Severin and P. Mosel and H. Merdji and A. Larue and S. Fröhlich and K.-A. Weber and U. Morgner and M. Kovacev and J. Biegert},
  title={\href{https://doi.org/10.1103/PRXQuantum.5.040319}{Evidence of the quantum-optical nature of high-harmonic generation}},
  journal={PRX Quantum},
  volume={5}, pages={040319},
  year={2024}
}

@article{stammer2024absence,
  title={\href{https://journals.aps.org/prresearch/abstract/10.1103/PhysRevResearch.6.L032033}{Absence of quantum optical coherence in high harmonic generation}},
  author={Stammer, Philipp},
  journal={Physical Review Research},
  volume={6},
  number={3},
  pages={L032033},
  year={2024},
  publisher={APS}
}

@article{stammer2026fluctuation,
  title={\href{https://arxiv.org/abs/2603.24377}{Fluctuation-induced symmetry breaking in high harmonic generation for bicircular quantum light}},
  author={Stammer, Philipp and Granados, Camilo and Rivera-Dean, Javier},
  journal={arXiv preprint arXiv:2603.24377},
  year={2026}
}

@article{stammer2026quantum,
  title={\href{https://journals.aps.org/pra/abstract/10.1103/w2lk-838z}{Quantum stochastic dynamics of nonlinear driven light emission}},
  author={Stammer, Philipp},
  journal={Phys. Rev. A},
  volume={114},
  number={1},
  pages={013712},
  year={2026},
  publisher={APS}
}

@article{stammer2026high,
  title={\href{https://iopscience.iop.org/article/10.1088/1361-6455/ae8630/meta}{High harmonic generation from a Bose-Einstein condensate}},
  author={Stammer, Philipp},
  journal={Journal of Physics B: Atomic, Molecular and Optical Physics},
  year={2026}
}

@article{stammer2024limitations,
  title={\href{https://www.nature.com}{On the limitations of the semi-classical picture in high harmonic generation}},
  author={Stammer, Philipp},
  journal={Nature Phys.},
  volume={20},
  number={7},
  pages={1040--1042},
  year={2024},
  publisher={Nature Publishing Group UK London}
}

@article{rivera2026attosecond,
  title={\href{https://iopscience.iop.org/article/10.1088/1361-6633/ae5847/meta}{Attosecond quantum optical interferometry}},
  author={Rivera-Dean, Javier and Petrovic, Lidija and Lewenstein, Maciej and Stammer, Philipp},
  journal={Reports on Progress in Physics},
  volume={89},
  number={4},
  pages={047901},
  year={2026},
  publisher={IOP Publishing}
}

@article{Maria_ArXiv_2024,
  title = {\href{https://arxiv.org/abs/2403.15337}{High Harmonic Generation by Bright Squeezed Vacuum}},
  journal = {Nature Phys.},
  volume = {20},
  pages = {1960--1965},
  author = {A. Rasputnyi and Z. Chen and M. Birk and O. Cohen and I. Kaminer and M. Krüger and D. Seletskiy and M. Chekhova and F. Tani},
  year = {2024}
}

@article{RSC24,
  author={J. Rivera-Dean and P. Stammer and M. F. Ciappina and M. Lewenstein},
  title={\href{https://doi.org/10.1103/4hdl-bdwj}{Structured Squeezed Light Allows for High-Harmonic Generation in Classical Forbidden Geometries}},
  journal={Phys. Rev. Lett.},
  volume={135}, pages={013801},
  year={2025}
}

@article{lewenstein2021generation,
  title={\href{https://www.nature.com/articles/s41567-021-01317-w}
  {Generation of optical Schr{\"o}dinger cat states in intense laser--matter interactions}},
  author={Lewenstein, Maciej and Ciappina, Marcelo F. and Pisanty, Emilio and Rivera-Dean, Javier and Stammer, Philipp and Lamprou, Th. and Tzallas, Paraskevas},
  journal={Nature Phys.},
  volume={17},
  number={10},
  pages={1104--1108},
  year={2021},
  publisher={Nature Publishing Group}
}

@article{hernandez2017extreme,
author = {Carlos Hern{\'a}ndez-Garc{\'i}a and Alex Turpin and Julio San Rom{\'a}n and Antonio Pic{\'o}n and Rokas Drevinskas and Ausra Cerkauskaite and Peter G. Kazansky and Charles G. Durfee and {\'I}{\~n}igo J. Sola},
title = {\href{https://opg.optica.org/optica/fulltext.cfm?uri=optica-4-5-520}{Extreme ultraviolet vector beams driven by infrared lasers}},
journal = {Optica},
volume = {4},
number = {5},
pages = {520--526},
year = {2017},
publisher = {Optical Society of America}

}

@article{otte2018spatial,
  author={E. Otte and K. Tekce and C. Denz},
  title={\href{https://doi.org/10.1088/2040-8986/aadef3}{Spatial multiplexing for tailored fully-structured light}},
  journal={J. Opt.},
  volume={20}, number={10}, pages={105606},
  year={2018}, publisher={IOP Publishing}
}

@article{fang2025ultrafast,
  author={Yiqi Fang and Zijian Lyu and Yunquan Liu},
  title={\href{https://www.nature.com/articles/s42254-025-00887-5}{Ultrafast physics with structured light}},
  journal={Nature Reviews Physics},
  volume={7}, number={12}, pages={713--727},
  year={2025}, publisher={Nature Publishing Group UK London}
}

@article{forbes2021structured,
  title={Structured light},
  author={Forbes, Andrew and De Oliveira, Michael and Dennis, Mark R},
  journal={Nature photonics},
  volume={15},
  number={4},
  pages={253--262},
  year={2021},
  publisher={Nature Publishing Group UK London}
}

@article{dorney2019controlling,
  author={Kevin M. Dorney and Laura Rego and Nathan J. Brooks and Julio San Rom{\'a}n and Chen-Ting Liao and Jennifer L. Ellis and Dmitriy Zusin and Christian Gentry and Quynh L. Nguyen and Justin M. Shaw and others},
  title={\href{https://www.nature.com/articles/s41566-018-0304-3}{Controlling the polarization and vortex charge of attosecond high-harmonic beams via simultaneous spin--orbit momentum conservation}},
  journal={Nature photonics},
  volume={13},
  number={2},
  pages={123--130},
  year={2019},
  publisher={Nature Publishing Group UK London}
}

@article{de2022extreme,
  author={Alba de las Heras and Alok Kumar Pandey and Julio San Rom{\'a}n and Javier Serrano and Elsa Baynard and Guillaume Dovillaire and Moana Pittman and Charles G. Durfee and Luis Plaja and Sophie Kazamias and others},
  title={\href{https://opg.optica.org/optica/fulltext.cfm?uri=optica-9-1-71}{Extreme-ultraviolet vector-vortex beams from high harmonic generation}},
  journal={Optica},
  volume={9}, number={1}, pages={71--79},
  year={2022}, publisher={Optical Society of America}
}

@article{geneaux2016synthesis,
  author={R. G{\'e}neaux and A. Camper and T. Auguste and O. Gobert and J. Caillat and R. Ta{\"i}eb and Thierry Ruchon},
  title={\href{https://www.nature.com/articles/ncomms12583}{Synthesis and characterization of attosecond light vortices in the extreme ultraviolet}},
  journal={Nat. Commun.},
  volume={7}, number={1}, pages={12583},
  year={2016}, publisher={Nature Publishing Group UK London}
}

@article{kaarcher2023generation,
  author={Victor K{\"a}rcher and Vyacheslav V. Kim and Andra Naresh Kumar Reddy and Helmut Zacharias and Rashid A. Ganeev},
  title={\href{https://pubs.acs.org/doi/10.1021/acsphotonics.3c01195}{Generation of complex vector and vortex extreme ultraviolet beams using the s-waveplate and spiral phase plate during high-order harmonics generation in argon}},
  journal={ACS Photonics},
  volume={10}, number={12}, pages={4519--4528},
  year={2023}, publisher={ACS Publications}
}

@article{dong2026generation,
  author={Jia-Hao Dong and Liang Xu and Qing-Qing Liang and Ji-Jun Feng and Song-Lin Zhuang and Yi Liu},
  title={\href{https://link.springer.com/article/10.1186/s43074-026-00260-4}{Generation and control of extreme-ultraviolet spatiotemporal skyrmions and vector hopfions with high harmonic generation}},
  journal={PhotoniX},
  volume={7}, number={1}, pages={40},
  year={2026}, publisher={Springer}
}

@article{schmidt2025self,
  author={David D. Schmidt and Jos{\'e} Miguel Pablos-Mar{\'\i}n and Cameron Clarke and Jonathan Barolak and Nathaniel Westlake and Alba de las Heras and Javier Serrano and Sergei Shevtsov and Peter Kazansky and Daniel Adams and others},
  title={\href{https://pubs.aip.org/aip/app/article/10/6/060801/3348214}{Self-interfering high harmonic beam arrays driven by Hermite--Gaussian beams}},
  journal={APL Photonics},
  volume={10}, number={6},
  year={2025}, publisher={AIP Publishing}
}

@article{turpin2017extreme,
  author={Alex Turpin and Laura Rego and Antonio Pic{\'o}n and Julio San Rom{\'a}n and Carlos Hern{\'a}ndez-Garc{\'\i}a},
  title={\href{https://www.nature.com/articles/srep43888}{Extreme ultraviolet fractional orbital angular momentum beams from high harmonic generation}},
  journal={Sci. Rep.},
  volume={7}, number={1}, pages={43888},
  year={2017}, publisher={Nature Publishing Group UK London}
}

@article{martin2025extreme,
  author={Rodrigo Mart{\'\i}n-Hern{\'a}ndez and Guan Gui and Luis Plaja and Henry C. Kapteyn and Margaret M. Murnane and Chen-Ting Liao and Miguel A. Porras and Carlos Hern{\'a}ndez-Garc{\'\i}a},
  title={\href{https://www.nature.com/articles/s41566-025-01699-w}{Extreme-ultraviolet spatiotemporal vortices via high harmonic generation}},
  journal={Nature Photonics},
  volume={19}, number={8}, pages={817--824},
  year={2025}, publisher={Nature Publishing Group UK London}
}

@article{camper2014high,
  author={A. Camper and T. Ruchon and D. Gauthier and O. Gobert and P. Sali{\`e}res and B. Carr{\'e} and T. Auguste},
  title={\href{https://journals.aps.org/pra/abstract/10.1103/PhysRevA.89.043843}{High-harmonic phase spectroscopy using a binary diffractive optical element}},
  journal={Phys. Rev. A},
  volume={89}, number={4}, pages={043843},
  year={2014}, publisher={APS}
}

@article{chini2009delay,
  author={Michael Chini and Hiroki Mashiko and He Wang and Shouyuan Chen and Chenxia Yun and Shane Scott and Steve Gilbertson and Zenghu Chang},
  title={\href{https://opg.optica.org/oe/fulltext.cfm?uri=oe-17-24-21459}{Delay control in attosecond pump-probe experiments}},
  journal={Opt. Express},
  volume={17}, number={24}, pages={21459--21464},
  year={2009}, publisher={Optical Society of America}
}

@article{csizmadia2025active,
  author={Tam{\'a}s Csizmadia and L{\'e}n{\'a}rd Guly{\'a}s Oldal and Barnab{\'a}s Gilicze and D{\'a}niel Kiss and Tam{\'a}s Bartyik and Katalin Varj{\'u} and Subhendu Kahaly and Bal{\'a}zs Major},
  title={\href{https://pubs.aip.org/aip/app/article/10/8/080803/3359808}{Active stabilization for ultralong acquisitions in an attosecond pump--probe beamline}},
  journal={APL Photonics},
  volume={10}, number={8},
  year={2025}, publisher={AIP Publishing}
}

@article{burger2017compact,
  author={Christian Burger and W. F. Frisch and T. M. Karda{\'s} and M. Trubetskov and Volodymyr Pervak and R. Moshammer and Boris Bergues and Matthias F. Kling and P. Wnuk},
  title={\href{https://opg.optica.org/oe/fulltext.cfm?uri=oe-25-25-31130}{Compact and flexible harmonic generator and three-color synthesizer for femtosecond coherent control and time-resolved studies}},
  journal={Opt. Express},
  volume={25}, number={25}, pages={31130--31139},
  year={2017}, publisher={Optical Society of America}
}

@article{raab2024highly,
  author={A.-K. Raab and Marvin Schmoll and Emma R. Simpson and Melvin Redon and Yuman Fang and Chen Guo and A.-L. Viotti and Cord L. Arnold and Anne L’Huillier and Johan Mauritsson},
  title={\href{https://pubs.aip.org/aip/rsi/article/95/7/073002/3303489}{Highly versatile, two-color setup for high-order harmonic generation using spatial light modulators}},
  journal={Review of Scientific Instruments},
  volume={95}, number={7},
  year={2024}, publisher={AIP Publishing}
}

@article{jansen2016spatially,
  author={G. S. M. Jansen and Denis Rudolf and Lars Freisem and K. S. E. Eikema and S. Witte},
  title={\href{https://opg.optica.org/optica/fulltext.cfm?uri=optica-3-10-1122}{Spatially resolved Fourier transform spectroscopy in the extreme ultraviolet}},
  journal={Optica},
  volume={3}, number={10}, pages={1122--1125},
  year={2016}, publisher={Optical Society of America}
}

@article{ardini2024generation,
  author={Benedetto Ardini and Fabian Richter and Lorenzo Uboldi and Paolo Cinquegrana and M. Danailov and Alexander Demidovich and Sarang Dev Ganeshamandiram and Sebastian Hartweg and Gabor Kurdi and Friedemann Landmesser and others},
  title={\href{https://iopscience.iop.org/article/10.1088/1361-6455/ad2e2d/meta}{Generation of interferometrically stable pulse pairs from a free-electron laser using a birefringent interferometer}},
  journal={Journal of Physics B: Atomic, Molecular and Optical Physics},
  volume={57}, number={7}, pages={075402},
  year={2024}, publisher={IOP Publishing}
}

@article{kuzkova2026attosecond,
  author={Nataliia Kuzkova and Pieter J. van Essen and Roy van der Linden and Brian de Keijzer and Rui E. F. Silva and {\'A}lvaro Jim{\'e}nez Gal{\'a}n and Peter M. Kraus},
  title={\href{https://www.science.org/doi/full/10.1126/sciadv.aeb4109}{Attosecond high-harmonic interferometry probes orbital- and band-dependent dipole phase in magnesium oxide}},
  journal={Science Advances},
  volume={12}, number={18}, pages={eaeb4109},
  year={2026}, publisher={American Association for the Advancement of Science}
}

@article{garcia2024topological,
  author={Ana Garc{\'\i}a-Cabrera and Roberto Boyero-Garc{\'\i}a and {\'O}scar Zurr{\'o}n-Cifuentes and Javier Serrano and Julio San Rom{\'a}n and Luis Plaja and Carlos Hern{\'a}ndez-Garc{\'\i}a},
  title={\href{https://www.nature.com/articles/s42005-023-01511-7}{Topological high-harmonic spectroscopy}},
  journal={Commun. Phys.},
  volume={7}, number={1}, pages={28},
  year={2024}, publisher={Nature Publishing Group UK London}
}

@article{watzel2020multipolar,
  author={Jonas W{\"a}tzel and Jamal Berakdar},
  title={\href{https://journals.aps.org/pra/abstract/10.1103/PhysRevA.101.043409}{Multipolar, polarization-shaped high-order harmonic generation by intense vector beams}},
  journal={Phys. Rev. A},
  volume={101}, number={4}, pages={043409},
  year={2020}, publisher={APS}
}

@article{das2026optical,
author = {Bikash K. Das and C. Granados and M.F. Ciappina},
title = {Optical vortices: revolutionizing the field of linear and nonlinear optics},
journal = {Adv. Phys.: X},
volume = {11},
number = {1},
pages = {2608076},
year = {2026},
doi = {10.1080/23746149.2025.2608076},
URL = {https://doi.org/10.1080/23746149.2025.2608076},
}

\setcounter{figure}{0}
\renewcommand{\thefigure}{S\arabic{figure}}
\renewcommand{\theHfigure}{supp.\arabic{figure}}
\section*{Supplementary Material}
\subsection{Structured driving fields}

\begin{figure*}[t]
\centering

\subfigure[]{
    \includegraphics[height=0.45\textwidth]{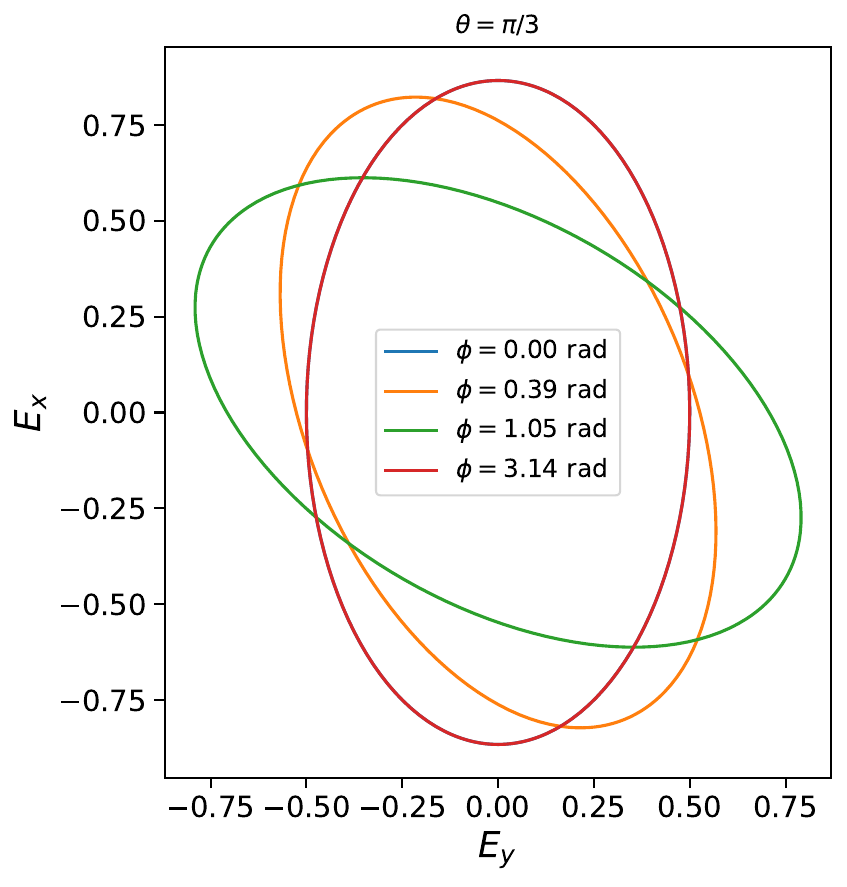}
    \label{fig:fre_field}
}
\hfill
\subfigure[]{
    \includegraphics[height=0.45\textwidth]{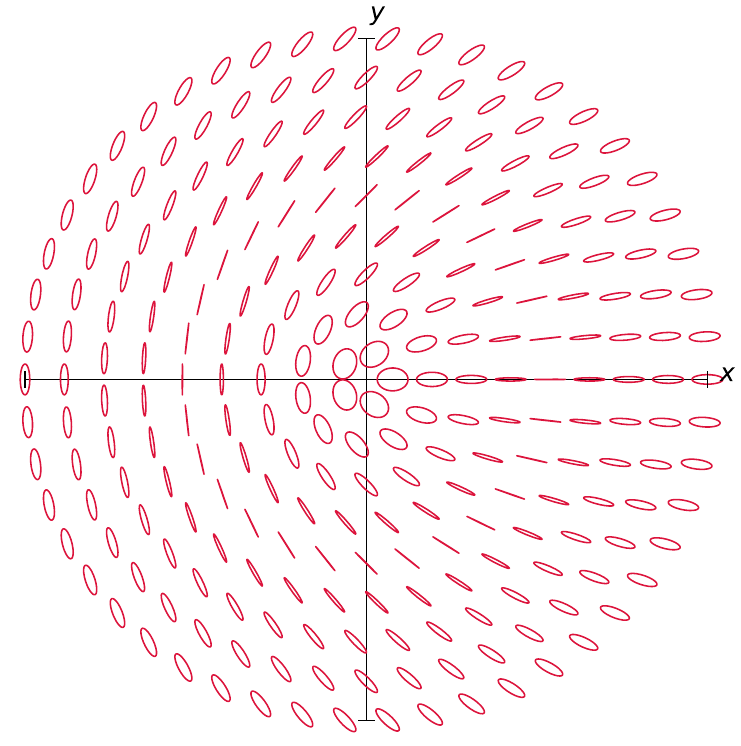}
    \label{fig:fpb_field}
}

\caption{Structured driving fields considered in this work.
(a) Family of rotating polarization ellipses (FRE) for fixed ellipticity
$\theta = \pi/3$. Variation of the parameter $\phi$ rotates the polarization
ellipse while preserving its shape and ellipticity.
(b) Full Poincaré beam (FPB) for the relative intensity parameter
$\theta = \pi/3$.
Parameters used: $\omega = 0.057$ a.u., $\alpha_x = 0.053$ a.u.,
$\delta = 0$, $\ell = 1$}.
\label{fig:structured_fields}
\end{figure*}

The two structured driving fields employed throughout this work are shown in Fig.~\ref{fig:structured_fields}. The family of rotating polarization ellipses (FRE) is characterized by the control parameters $(\theta,\phi)$, where $\theta$ determines the ellipticity and $\phi$ specifies the orientation of the polarization ellipse. For a fixed value of $\theta$, varying $\phi$ continuously rotates the polarization ellipse in the transverse plane while preserving its shape and ellipticity, thereby generating the closed parameter-space trajectories investigated in the main text (Fig.~\ref{fig:structured_fields}(a)).

The second driving configuration is the full Poincaré beam (FPB), which is formed from the coherent superposition of the Gaussian and vortex modes. As a result, the beam exhibits a spatially varying polarization distribution across its transverse profile, providing a structured driving field for the HHG process (Fig.~\ref{fig:structured_fields}(b)).
\subsection{Numerical calculation of the HHG spectra}
The high-harmonic spectra shown in Fig.~\ref{fig:hhg_response}(a) were obtained by numerically evaluating the time-dependent dipole moment within the strong-field approximation (SFA), as given by Eq.~(\ref{eq7}). For each value of the ellipticity parameter, the driving electric field of Eq.~(\ref{eq9}) was used to calculate the induced dipole moment over the entire laser pulse. The calculations were performed using a laser pulse with carrier frequency $\omega=0.057$ a.u. ($\lambda=800 $ nm) and peak electric-field amplitude $E_0=0.053$ a.u. (corresponding to a laser intensity $I=1\times10^{14}$ W/cm$^2$). A pulse envelope of the form $f(t)=\sin^2(\omega t/2n_{\rm cy})$ was employed, and the dipole moment was evaluated over the interval $0\le t\le n_{\rm cy}2\pi/\omega$ with $n_{\rm cy}=5$, corresponding to a pulse duration of approximately $13.3$ fs. A hydrogen atom with ionization potential $I_p=0.5$ a.u. was employed as a target.

No explicit trajectory selection was applied in the present calculations; therefore, both the short and long electron trajectories contribute to the calculated HHG spectra. After obtaining the time-dependent dipole components $D_x(t)$ and $D_y(t)$ from Eq.~(\ref{eq7}), the harmonic spectrum was calculated by Fourier transforming the dipole signal. The total HHG spectrum was evaluated as the sum of the spectral intensities of the two orthogonal polarization components.

Similarly, the HHG response for the full Poincar\'e beam (FPB), shown in Fig.~\ref{fig:hhg_response}(b), was calculated using the structured driving field defined in Eq.~(\ref{eq17}). As illustrated in Fig.~\ref{fig:structured_fields}, the coherent superposition of the Gaussian and vortex modes produces a spatially varying polarization distribution across the transverse beam profile. Consequently, the local polarization state, and hence the driving-field ellipticity, changes from one spatial point to another. The SFA dipole response of Eq.~(\ref{eq7}) was therefore evaluated independently at each transverse position $(\rho,\phi)$ using the corresponding local driving field. The harmonic spectrum was then obtained by Fourier transforming the local time-dependent dipole moment, yielding a spatially resolved HHG response. The intensity distribution of the selected harmonic order shown in Fig.~\ref{fig:hhg_response}(b) is obtained by repeating this procedure over the entire transverse beam profile.
\subsection{High-harmonic response}

\begin{figure*}[t]
\centering

\subfigure[]{
    \includegraphics[height=0.3\textwidth]{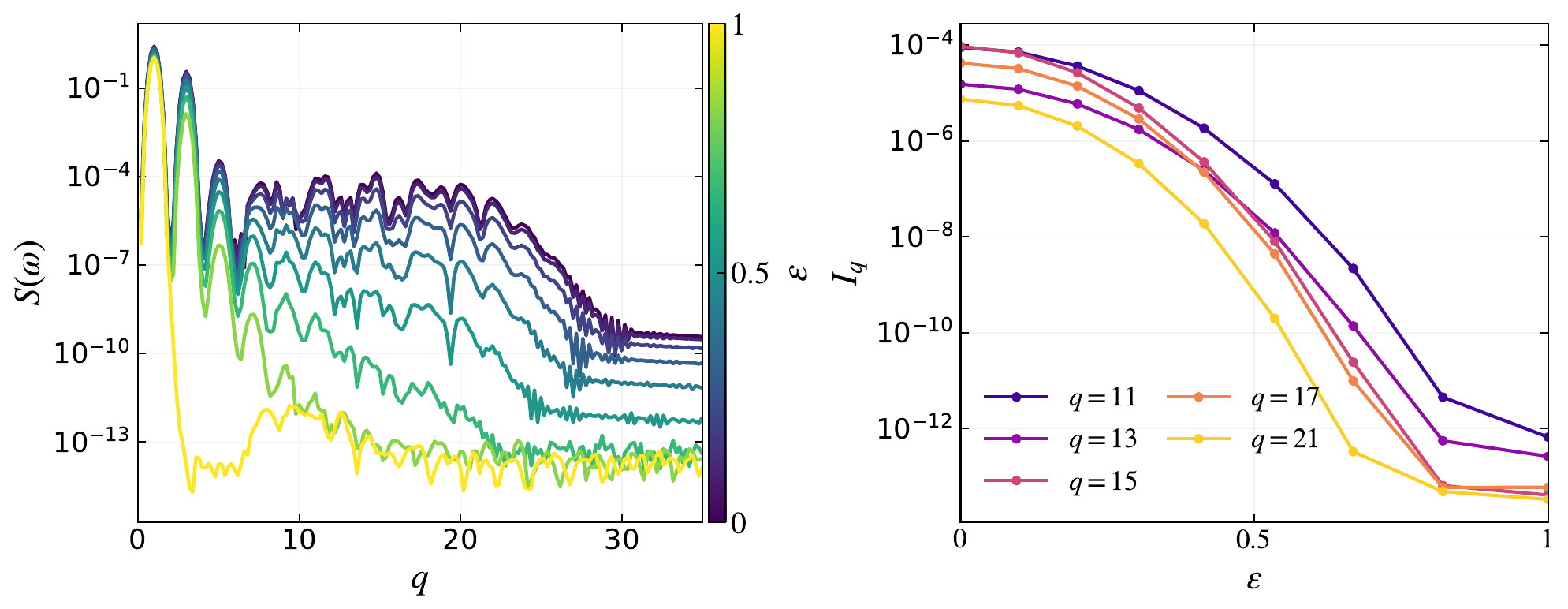}
    \label{fig:fre_hhg}
}
\hfill
\subfigure[]{
    \includegraphics[height=0.5\textwidth]{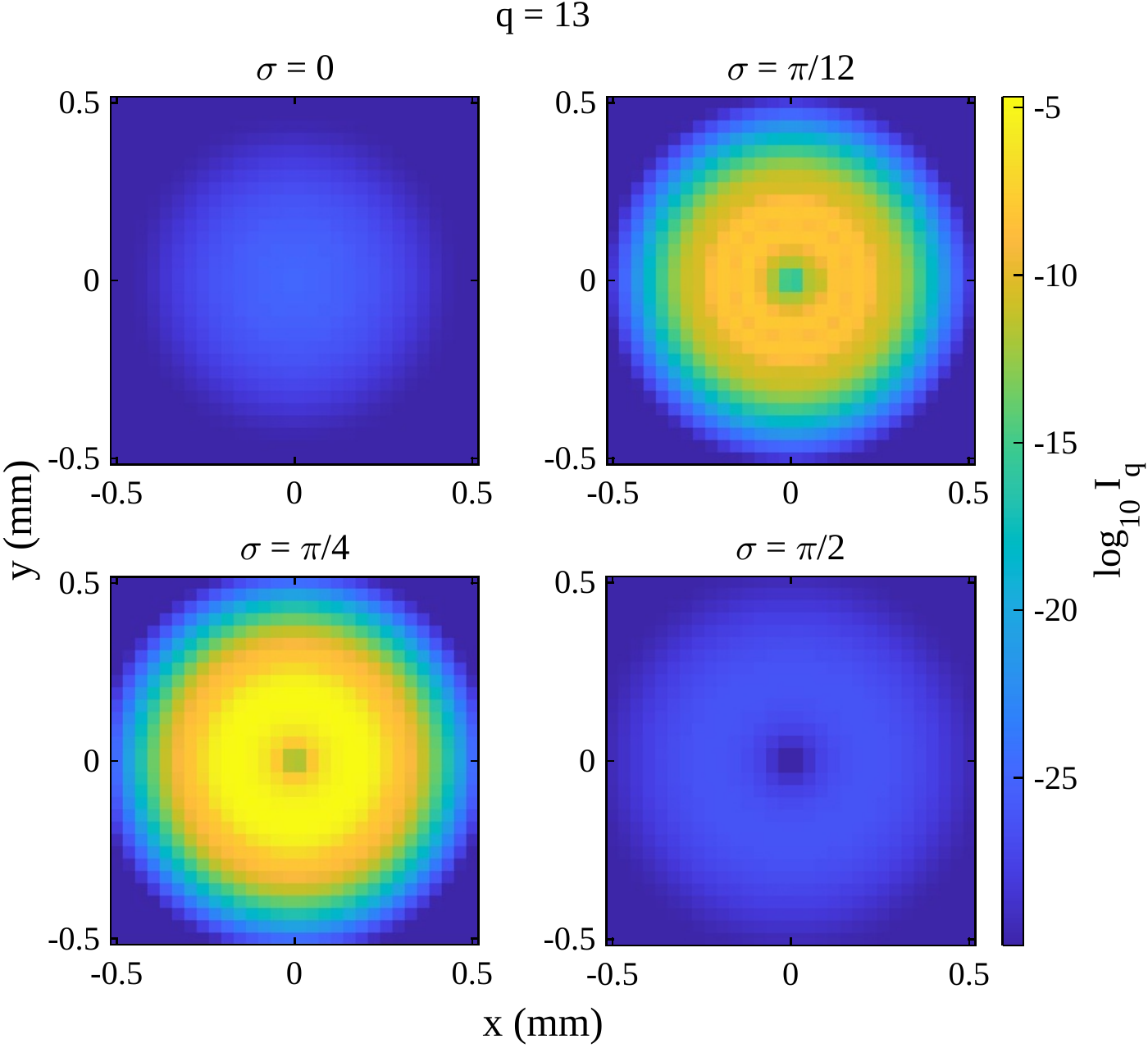}
    \label{fig:fpb_hhg}
}

\caption{High-harmonic response generated by the structured driving fields.
(a) High-harmonic spectra obtained for the family of rotating polarization
ellipses for different ellipticities. Parameters used:
$\omega = 0.057$ a.u., $\alpha_x = 0.053$ a.u., $I_p = 0.5$ a.u.,
$n_{\mathrm{cy}} = 5$ (corresponding to a pulse duration
$\tau_p = n_{\mathrm{cy}}2\pi/\omega \approx 13.3$ fs), and
$\alpha = 0.8I_p$.
(b) Spatial distribution of the 13th-harmonic intensity generated under
full Poincaré beam driving for different values of the parameter $\theta$.
The parameter $\theta$ controls the relative intensity contribution of the
Gaussian and vortex components of the driving field.
Parameters used: $E_0 = 0.5$ a.u., $\omega = 0.057$ a.u., $I_p = 0.5$ a.u.,
$w_0 = 0.3125~\mathrm{mm}$ (millimeter)
$\rho = 4\omega_0$ and $n_{\mathrm{cyc}} = 5$.}
\label{fig:hhg_response}
\end{figure*}

The corresponding high-harmonic response generated by the two structured driving fields is presented in Fig.~\ref{fig:hhg_response}. The HHG spectra obtained for the family of rotating polarization ellipses are shown in Fig.~\ref{fig:hhg_response}(a). As the ellipticity increases, the harmonic yield decreases, consistent with the well-known suppression of HHG driven by elliptically polarized fields. The right panel further shows the intensity of selected harmonic orders as a function of ellipticity, demonstrating a significant reduction in harmonic intensity with increasing ellipticity.

The spatial response obtained under full Poincaré beam driving is shown in Fig.~\ref{fig:hhg_response}(b). The parameter $\theta$ controls the relative intensity contribution of the Gaussian and vortex components, leading to the characteristic evolution of the spatial distribution of the 13th-harmonic intensity across the beam profile.

\end{document}